\documentclass[runningheads,a4paper]{llncs}

\usepackage{fullpage}

\usepackage{amssymb}      
\usepackage{amsmath}      
\usepackage{array}        
\usepackage{caption}
\usepackage{hyperref}
\usepackage{graphicx}
\usepackage{listings}
\usepackage{multirow}
\usepackage{pifont}       
\usepackage{rotating}
\usepackage{float}        
\usepackage{subfig}       
\usepackage{tikz} \usetikzlibrary{chains,positioning,scopes} 
\usepackage[misc]{ifsym}  

\usepackage{marvosym}     
\usepackage{xcolor}

\newcommand{\SI}{\ding{51}}
\newcommand{\NO}{\ding{55}}
\newcommand{\cG}{{\cal G}}

\colorlet{punct}{red!60!black}
\definecolor{background}{HTML}{FFFFFF}
\definecolor{delim}{RGB}{20,105,176}
\colorlet{numb}{magenta!60!black}

\lstdefinelanguage{json}{
    basicstyle=\normalfont\ttfamily,
    numbers=left,
    numberstyle=\scriptsize,
    stepnumber=1,
    numbersep=8pt,
    showstringspaces=false,
    breaklines=true,
    frame=lines,
    backgroundcolor=\color{background},
    literate=
     *{0}{{{\color{numb}0}}}{1}
      {1}{{{\color{numb}1}}}{1}
      {2}{{{\color{numb}2}}}{1}
      {3}{{{\color{numb}3}}}{1}
      {4}{{{\color{numb}4}}}{1}
      {5}{{{\color{numb}5}}}{1}
      {6}{{{\color{numb}6}}}{1}
      {7}{{{\color{numb}7}}}{1}
      {8}{{{\color{numb}8}}}{1}
      {9}{{{\color{numb}9}}}{1}
      {:}{{{\color{punct}{:}}}}{1}
      {,}{{{\color{punct}{,}}}}{1}
      {\{}{{{\color{delim}{\{}}}}{1}
      {\}}{{{\color{delim}{\}}}}}{1}
      {[}{{{\color{delim}{[}}}}{1}
      {]}{{{\color{delim}{]}}}}{1},
}

\begin{document}

\mainmatter

\title{Measuring user influence on Twitter: A survey\thanks{\copyright Published version: http://dx.doi.org/10.1016/j.ipm.2016.04.003}\thanks{\copyright 2016. This manuscript version is made available under the CC-BY-NC-ND 4.0 license http://creativecommons.org/licenses/by-nc-nd/4.0/
}}
\author{Fabi\'an Riquelme \and Pablo Gonz\'alez-Cantergiani}

\institute{CITIAPS - Universidad de Santiago, Chile.\\
	\email{$\{$fabian.riquelme.c, pablo.gonzalezca$\}$@usach.cl}
}

\date{}

\maketitle

\begin{abstract}
Centrality is one of the most studied concepts in social network analysis. There is a huge literature regarding centrality measures, as ways to identify the most relevant users in a social network. The challenge is to find measures that can be computed efficiently, and that can be able to classify the users according to relevance criteria as close as possible to reality.
We address this problem in the context of the Twitter network, an online social networking service with millions of users and an impressive flow of messages that are published and spread daily by interactions between users. Twitter has different types of users, but the greatest utility lies in finding the most influential ones.
The purpose of this article is to collect and classify the different Twitter influence measures that exist so far in literature. These measures are very diverse. Some are based on simple metrics provided by the Twitter API, while others are based on complex mathematical models. Several measures are based on the PageRank algorithm, traditionally used to rank the websites on the Internet. Some others consider the timeline of publication, others the content of the messages, some are focused on specific topics, and others try to make predictions. We consider all these aspects, and some additional ones.
Furthermore, we include measures of activity and popularity, the traditional mechanisms to correlate measures, and some important aspects of computational complexity for this particular context.

\end{abstract}

\section{Introduction}

Since the origins of social network analysis, there has been interest in identifying the most relevant actors of a social network. With the rise of the Internet and technology, online social networks (OSNs) turned into challenging cases of study, on which big data content and complex interpersonal ties among actors converge. Knowing the influence of users and being able to predict it can be useful for many applications, such as viral marketing~\cite{DR01,KKT03,RD02}, information propagation~\cite{GLGT04,GH06}, searching~\cite{AA05}, expertise recommendation~\cite{STLS06}, social customer relationship management~\cite{LPLSLX14}, percolation theory~\cite{MM15}, etc.

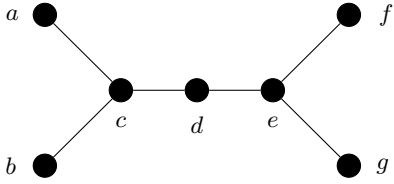
\begin{figure}[t]
\begin{center}
\begin{tikzpicture}[every node/.style={circle}, node distance=20mm, >=latex]
\node[](b) at (0,2)[draw,fill,label=left: $a$] {};
\node[](a) at (0,0)[draw,fill,label=left: $b$] {};
\node[](c) at (1,1)[draw,fill,label=below:$c$] {};
\node[](d) at (2,1)[draw,fill,label=below:$d$] {};
\node[](e) at (3,1)[draw,fill,label=below:$e$] {};
\node[](g) at (4,2)[draw,fill,label=right:$f$] {};
\node[](f) at (4,0)[draw,fill,label=right:$g$] {};
\draw[-] (a) to node {}(c);
\draw[-] (b) to node {}(c);
\draw[-] (c) to node {}(d);
\draw[-] (d) to node {}(e);
\draw[-] (e) to node {}(f);
\draw[-] (e) to node {}(g);
\end{tikzpicture}
\end{center}
\caption{An undirected graph representing interpersonal ties (edges) between users (nodes).\label{fig:example}}
\end{figure}

A social network can be represented by a graph, as in Figure~\ref{fig:example}, whose nodes can represent the users, and the edges the interpersonal ties among them. The {\em centrality} of a node refers to its relative importance within the network to which it belongs. There is a vast literature about centrality measures to identify the most important actors on a social network~\cite{WF94,ST11}. Each measure is based on different relevance criteria. For example, two of the most traditional measures are {\em degree} and {\em closeness}~\cite{Fre79}. The first one considers for each node the number of its adjacent edges, while the second one considers the minimum sum of the shortest paths from a node to all the other nodes within the network. In Figure~\ref{fig:example} nodes $c$ and $e$ have the higher degree, with three adjacent edges, followed by node $d$, with two adjacent edges. However, for the same graph, the higher closeness is for node $d$, with a sum of shortest paths equals $10$, followed by nodes $c$ and $e$, with a sum equals $11$. For both measures, nodes $a$, $b$, $f$ and $g$ are the last in the ranking. 
Depending on the topology of the network, some other traditional measures are {\em betweenness}~\cite{Fre79} and {\em eigenvector}~\cite{Bon72}. There are also centrality measures based on how much information can be dispersed through the nodes of a network~\cite{FBW91,GFE13}. Such measures, as well as other derivatives, can also be used in the context of OSNs like Twitter, although do not take into account all the features of these particular services~\cite{NVMR15}.

Besides the centrality of a node, it is important to know how a small set of initially active nodes can propagate some ``germs'' (ideas, trends, fashions, rules, ambitions, etc.) to other distant nodes in the network. For instance, in Figure~\ref{fig:example}, node $f$ can only be directly influenced by node $e$, but node $b$ could also propagate his influence until $f$, in a cascade behaviour through the nodes $c$, $d$ and $e$. This phenomenon, known as the {\em spread of influence}, has been empirically verified on social networks several times~\cite{ZHL12}. Furthermore, there exist centrality measures explicitly based on this phenomenon~\cite{MRS14}.

Only in Twitter, spread of influence has been used for different purposes, such as political sciences~\cite{MRDSF14}, human mobility~\cite{GBRM11}, transportation~\cite{LLKC12}, rumor spreading~\cite{BM12}, epidemiology~\cite{SVTG14}, among many others. Twitter\footnote{\url{www.twitter.com}} was created in 2006, and since 2010 is being studied extensively in the contexts of social network analysis, computer science, and sociology. One of the most studied problems in Twitter is the identification of influential users. This problem is especially important, considering the high percentage of users who are often inactive or do not provide additional information~\cite{AK11}. Furthermore, the criteria for identifying influential users are as many as the growing number of techniques to rank them.

In this article we review the numerous investigations that have been carried out regarding the definition of influence measures on Twitter. Although the earliest studies date just from 2009, the amount of literature on the subject is extensive. Table~\ref{tab:measures} classifies the studied measures under different criteria. First column splits the measures by activity, popularity and influence criteria. The reason for this classification is explained in Section~\ref{sec:inf-user}. The second column presents each measure and its corresponding reference. Note that {\em General activity} is the only new measure proposed in this paper. In the third group of columns we mark as yes (\SI) or no (\NO) if the measures use metrics based on follow-up relationships (F), retweets (RT), mentions (M), replies (RP), favorites or likes (FT). All these kind of Twitter relationships are explained in Section~\ref{sec:twitter}. Furthermore, we distinguish those measures that are based on the PageRank algorithm, those that consider timeline, and those that require content analysis. The latter column shows the computational complexity for each measure. This column is explained later in Section~\ref{sec:comcom}.

\begin{table}[t!]
\begin{center}
{\scriptsize
\begin{tabular}{c|c|lr|c|c|c|c|c|c|c|c|c}
\multicolumn{1}{c}{} 
  & & & & \multicolumn{5}{c|}{Kind of metric used}& & & Content & Time\\\cline{5-9}
  \multicolumn{2}{c|}{ } & Name & Ref. & F & RT & M & RP & FT & PageRank & Timeline & analysis & complexity\\\hline
\multicolumn{2}{c|}{\multirow{8}{*}{\rotatebox{90}{Activity}}}
  &Tweet count score     &\cite{NRXT13}  &\NO &\SI &\NO &\NO &\NO & \NO &\NO &\NO &$O(T)$ \\[-.5ex]
\multicolumn{1}{c}{}
& &General activity      &               &\NO &\SI &\SI &\SI &\SI & \NO &\NO &\NO &$O(T)$ \\[-.5ex]
\multicolumn{1}{c}{}
& &Topical signal        &\cite{PC11}    &\NO &\SI &\SI &\SI &\NO & \NO &\NO &\SI &$\Omega(T)$ \\[-.5ex]
\multicolumn{1}{c}{}
& &Signal strength       &\cite{PC11}    &\NO &\SI &\NO &\SI &\NO & \NO &\NO &\NO &$O(T)$ \\[-.5ex]
\multicolumn{1}{c}{}
& &Effective readers     &\cite{LKPM10}  &\SI &\NO &\NO &\SI &\NO & \NO &\SI &\SI &limited request\\[-.5ex]
\multicolumn{1}{c}{}
& &ActivityScore         &\cite{YLLH13}  &\SI &\NO &\NO &\SI &\NO & \NO &\SI &\NO &limited request\\[-.5ex]
\multicolumn{1}{c}{}
& &DiscussRank           &\cite{JTB12}   &\SI &\SI &\SI &\NO &\NO & \NO &\NO &\NO &limited request\\[-.5ex]
\multicolumn{1}{c}{}
& &Competency            &\cite{AKRO15}  &\NO &\NO &\NO &\NO &\NO & \NO &\NO &\SI &? (high)\\[-.5ex]
\multicolumn{1}{c}{}
& &IP Influence          &\cite{RGAH11}  &\SI &\SI &\NO &\NO &\NO & \NO &\SI &\SI &limited request\\\hline
\multicolumn{2}{c|}{\multirow{8}{*}{\rotatebox{90}{Popularity}}}
  &FollowerRank          &\cite{NTC10}   &\SI &\NO &\NO &\NO &\NO & \NO &\NO &\NO &$O(1)$ \\[-.5ex]
\multicolumn{1}{c}{}
& &Popularity            &\cite{AKRO15}  &\SI &\NO &\NO &\NO &\NO & \NO &\NO &\NO &$O(1)$ \\[-.5ex]
\multicolumn{1}{c}{}
& &Paradoxical discounted&\cite{Gay13}   &\SI &\NO &\NO &\NO &\NO & \NO &\NO &\NO &limited request\\[-.5ex]
\multicolumn{1}{c}{}
& &Network Score         &\cite{PC11}    &\SI &\NO &\NO &\NO &\NO & \NO &\SI &\SI &limited request\\[-.5ex]
\multicolumn{1}{c}{}
& &$A$ Score             &\cite{SST13}   &\SI &\SI &\SI &\SI &\NO & \NO &\NO &\NO &$O(T\cdot k)$\\[-.5ex]
\multicolumn{1}{c}{}
& &$AA$ Score            &\cite{SST13}   &\SI &\SI &\SI &\SI &\NO & \NO &\NO &\NO &$O(T\cdot k^2)$\\[-.5ex]
\multicolumn{1}{c}{}
& &$AAI$ Score           &\cite{SST13}   &\SI &\SI &\SI &\SI &\NO & \NO &\NO &\NO &limited request\\[-.5ex]
\multicolumn{1}{c}{}
& &Action-Reaction       &\cite{SST14}   &\SI &\SI &\SI &\SI &\NO & \NO &\NO &\NO &$O(T\cdot k^2)$ \\[-.5ex]
\multicolumn{1}{c}{}
& &Starrank              &\cite{KC10}    &\NO &\NO &\SI &\NO &\NO & \SI &\SI &\NO &$O(T+n^{2.3727})$\\\hline
\multirow{49}{*}{\rotatebox{90}{Influence measures}}
& \multirow{4}{*}{\rotatebox{90}{General}}
  &Closeness             &\cite{HW11}    &\SI &\NO &\NO &\NO &\NO & \NO &\NO &\NO &limited request\\[-.5ex]
& &Betweenness           &\cite{JW13}    &\SI &\NO &\NO &\NO &\NO & \NO &\NO &\NO &limited request\\[-.5ex]
& &$H$-index             &\cite{Hir10}   &\NO &\SI &\NO &\SI &\SI & \NO &\NO &\NO &$O(T+n)$\\[-.5ex]
& &Velocity              &\cite{GBFFG11} &\SI &\NO &\SI &\NO &\NO & \NO &\SI &\NO &$O(T)$\\\cline{3-13}
& \multirow{17}{*}{\rotatebox{90}{Based on metrics or PageRank}}
  &Retweet Impact        &\cite{PC11}    &\NO &\SI &\NO &\NO &\NO & \NO &\NO &\NO &$O(T\cdot k)$\\[-.5ex]
& &Mention Impact        &\cite{PC11}    &\NO &\NO &\SI &\NO &\NO & \NO &\NO &\NO &$O(T\cdot k)$\\[-.5ex]
& &SNP                   &\cite{AK11}    &\SI &\SI &\SI &\SI &\NO & \NO &\NO &\NO &$O(T\cdot k)$\\[-.5ex]
& &Content \& conversation&\cite{HBV11}  &\SI &\SI &\SI &\SI &\NO & \NO &\NO &\NO &limited request\\[-.5ex]
& &IP Influence          &\cite{RGAH11}  &\SI &\SI &\NO &\NO &\NO & \NO &\SI &\SI &limited request\\[-.5ex]
& &TunkRank              &\cite{Tun09}   &\SI &\SI &\NO &\NO &\NO & \SI &\NO &\NO &limited request\\[-.5ex]
& &TrueTop               &\cite{ZZSZZ15} &\NO &\SI &\SI &\SI &\NO & \NO &\NO &\NO &$O(T+n^{2.3727})$\\[-.5ex]
& &UserRank              &\cite{MS12}    &\SI &\NO &\NO &\NO &\NO & \SI &\NO &\NO &limited request\\[-.5ex]
& &DIS                   &\cite{HLLC13}  &\SI &\SI &\NO &\NO &\NO & \SI &\NO &\NO &limited request\\[-.5ex]
& &Influence Rank        &\cite{HW11}    &\SI &\SI &\SI &\NO &\SI & \SI &\NO &\NO &limited request\\[-.5ex]
& &By similarity factor  &\cite{LCCJ13}  &\SI &\SI &\SI &\SI &\NO & \SI &\NO &\SI &limited request\\[-.5ex]
& &InfRank               &\cite{JTB12}   &\NO &\SI &\NO &\NO &\NO & \SI &\NO &\NO &limited request\\[-.5ex]
& &LeadRank              &\cite{JTB12}   &\NO &\SI &\SI &\NO &\NO & \NO &\NO &\NO &limited request\\[-.5ex]
& &SpreadRank            &\cite{DJZHHZ13}&\NO &\SI &\NO &\NO &\NO & \SI &\SI &\NO &$O(T+n^{2.3727})$\\[-.5ex]
& &ProfileRank           &\cite{SGMZ13}  &\SI &\SI &\SI &\SI &\NO & \SI &\SI &\NO &$O(T+n^{2.3727})$\\[-.5ex]
& &MultiRank             &\cite{DJZH13}  &\NO &\SI &\NO &\SI &\NO & \SI &\SI &\SI &? (high)\\[-.5ex]
& &TURank                &\cite{YTAK10}  &\SI &\SI &\NO &\NO &\NO & \SI &\NO &\NO &limited request\\[-.5ex]
& &By TAC                &\cite{LWH13}   &\SI &\SI &\NO &\SI &\NO & \SI &\SI &\NO &limited request\\\cline{3-13}
& \multirow{23}{*}{\rotatebox{90}{Topical-sensitive}}
  &Alpha centrality      &\cite{OGPJ13}  &\NO &\SI &\NO &\NO &\NO & \NO &\SI &\SI &$\Omega(T+n^{2.3727})$\\[-.5ex]
& &$T$-index             &\cite{KRGVBD13}&\NO &\SI &\NO &\NO &\NO & \NO &\NO &\SI &$\Omega(T+n)$\\[-.5ex]
& &Information Diffusion &\cite{PC11}    &\SI &\NO &\NO &\NO &\NO & \NO &\SI &\SI &limited request\\[-.5ex]
& &Topic-Specific Author &\cite{KF11}    &\SI &\SI &\SI &\NO &\NO & \NO &\NO &\SI &limited request\\[-.5ex]
& &By effective audience &\cite{SZL13}   &\SI &\SI &\NO &\SI &\NO & \NO &\SI &\SI &limited request\\[-.5ex]
& &TRank                 &\cite{MF15}    &\SI &\SI &\NO &\NO &\SI & \NO &\NO &\SI &limited request\\[-.5ex]
& &RetweetRank           &\cite{XNT14}   &\NO &\SI &\NO &\NO &\NO & \SI &\NO &\SI &$\Omega(T+n^{2.3727})$\\[-.5ex]
& &MentionRank           &\cite{XNT14}   &\NO &\NO &\SI &\NO &\NO & \SI &\NO &\SI &$\Omega(T+n^{2.3727})$\\[-.5ex]
& &TwitterRank           &\cite{WLJH10}  &\SI &\NO &\NO &\NO &\NO & \SI &\NO &\SI &limited request\\[-.5ex]
& &InterRank             &\cite{SML13}   &\SI &\NO &\NO &\NO &\NO & \SI &\NO &\SI &limited request\\[-.5ex]
& &Topic-Entity PageRank &\cite{CMC14}   &\NO &\SI &\NO &\NO &\NO & \SI &\NO &\SI &$\Omega(T+n^{2.3727})$\\[-.5ex]
& &TIURank               &\cite{LSML14}  &\NO &\SI &\NO &\NO &\NO & \SI &\NO &\SI &limited request\\[-.5ex]
& &ARI                   &\cite{HMR14}   &\NO &\SI &\SI &\SI &\NO & \SI &\NO &\SI &$\Omega(T+n^{2.3727})$\\[-.5ex]
& &Twitter user rank     &\cite{NRXT13}  &\NO &\SI &\SI &\SI &\NO & \SI &\NO &\SI &$\Omega(T+n^{2.3727})$\\[-.5ex]
& &TS-SRW                &\cite{KVP15}   &\NO &\NO &\SI &\NO &\NO & \SI &\NO &\SI &$\Omega(T+n^{2.3727})$\\[-.5ex]
& &Topical Authority     &\cite{HFG13}   &\NO &\SI &\NO &\NO &\NO & \SI &\NO &\SI &$\Omega(T+n^{2.3727})$\\[-.5ex]
& &IARank                &\cite{CS12}    &\SI &\SI &\SI &\SI &\NO & \NO &\NO &\SI &$\Omega(T\cdot k^2)$\\[-.5ex]
& &SNI                   &\cite{HCA14}   &\SI &\SI &\SI &\NO &\NO & \SI &\SI &\SI &$\Omega(T+n^{2.3727})$\\[-.5ex]
& &By polarity \& others &\cite{BCMGA12} &\SI &\SI &\SI &\SI &\NO & \NO &\NO &\SI &limited request\\[-.5ex]
& &By sucesptibility...  &\cite{LL15}    &\SI &\SI &\NO &\NO &\NO & \NO &\NO &\SI &limited request\\[-.5ex]
& &By tweets graph       &\cite{SN13}    &\SI &\SI &\NO &\SI &\NO & \NO &\NO &\SI &? (high)\\[-.5ex]
& &WRA                   &\cite{YLLH13}  &\SI &\SI &\NO &\SI &\NO & \NO &\SI &\SI &limited request\\[-.5ex]
& &FLDA                  &\cite{BTSBC14} &\SI &\NO &\NO &\NO &\NO & \NO &\NO &\SI &? (high)\\[-.5ex]
& &Leadership            &\cite{AKRO15}  &\SI &\NO &\NO &\NO &\NO & \NO &\NO &\SI &? (high)\\[-.5ex]\cline{3-13}
& \multirow{7}{*}{\rotatebox{90}{Predictive}}
  &AWI model             &\cite{YZ12}    &\NO &\SI &\NO &\NO &\NO & \NO &\SI &\NO &?\\[-.5ex]
& &ACQR Framework        &\cite{CXZW13}  &\NO &\SI &\NO &\SI &\NO & \NO &\SI &\SI &?\\[-.5ex]
& &TNIM                  &\cite{DJZZHY13}&\SI &\NO &\NO &\NO &\NO & \NO &\SI &\SI &limited request\\[-.5ex]
& &Author Ranking        &\cite{VRSJLR14}&\SI &\NO &\NO &\NO &\NO & \NO &\SI &\SI &?\\[-.5ex]
& &ReachBuzzRank         &\cite{SVH13}   &\SI &\SI &\SI &\SI &\NO & \SI &\SI &\SI &?\\[-.5ex]
& &IDM-CTMP              &\cite{LPLSLX14}&\NO &\SI &\SI &\SI &\NO & \NO &\SI &\SI &?\\[-.5ex]
& &Parameterless mixture &\cite{BR15}    &\SI &\SI &\SI &\NO &\NO & \NO &\SI &\NO &?\\
\end{tabular}
\caption{Classification of measures in order of appearance in the article. F refers to follow-up relationships, RT to retweets, M to mentions, RP to replies, and FT to favorite or liked tweets. It is also indicated if the measure uses the PageRank algorithm, timeline, or topical analysis. In the latter column, $T$ is the number of tweets, $n$ the number of users, and $k$ the length of an auxiliary vector (see Section~\ref{sec:comcom}).}\label{tab:measures}
}
\end{center}
\end{table}

As far as we know, this is the first attempt to bring together all centrality measures used in the Twitter network. However, there are also other interesting partial surveys in the area. For example, in 2013, Gayo-Avello~\cite{Gay13} explained a dozen frequently used measures. These measures have been included here, along with other more than fifty. In a more recent article, Kardara et al.~\cite{KPPTV15} also mention some few metrics and measures. However, their main contribution is to propose a two-dimensional taxonomy to classify centrality measures: On the one hand, the authors classify the measures at {\em global}, {\em local}, or {\em glocal}, where {\em local} is our topical-sensitive criterion (see Table~\ref{tab:measures}); on the other hand, they classify the measures in {\em graphical} (those based on topological criteria), {\em contentual} (those based on content analysis) and {\em holistic} (an hybrid of the previous two). That classification is not as complete as the one given for Twitter in this paper, but is also suitable for non-Twitter measures. More common are the surveys related with centrality measures for general OSNs. Probst et al.~\cite{PGP13} try to classify different types of centrality criteria. In addition to our criteria of activity (``how active one is'') and popularity (``how known one is''), they consider other two: ``who one is'' and ``what one knows''. This classification is not clear enough to distinguish the influence criterion, which is our main interest. Vogiatzis~\cite{Vog13} also considers a few centrality measures aimed at different OSNs. Regarding Twitter, the author focuses on the same kind of relationships we consider: retweets, mentions, replies, etc. Finally, Sun and Tang~\cite{ST11} make a general summary about the social influence analysis for social networks. Regarding centrality measures, they only mention the most traditional ones, such as degree, closeness and betweenness.

In the next section we present the main concepts, characteristics and metrics regarding the Twitter network. In Section~\ref{sec:API} we explain how to obtain the data through the Twitter API, and we present the anatomy of a tweet. In Section~\ref{sec:inf-user} we talk about the ambiguity of the influential user's definition, and we divide the different measures according to user activity, popularity and influence. Section~\ref{sec:activity} is devoted to activity measures, Section~\ref{sec:popularity} to popularity measures, and Section~\ref{sec:influence} to influence measures. In Section~\ref{sec:add-aspects} we address the computational complexity of the measures, as well as some ways to correlate them with each other. Here we also introduce a case study in order to illustrate the variation of results for different centrality criteria. In Section~\ref{sec:open} we mention the main open problems related to this topic. We finish this work with some conclusions.

\section{A brief description of the Twitter network}
\label{sec:twitter}

Nowadays, Twitter is one of the main online social networking services on the world.\footnote{There are many other microblog systems on the web. After Twitter, the most popular is the chinese Sina Weibo: \url{www.weibo.com}.} It is a microblog service where users can send and read short 140-character messages called {\em tweets}. Tweets may include plain text, URL pages, images, mentions to other users (preceded by the symbol ``@'') and {\em hashtags}, which are words that one decide to highlight by placing the symbol ``$\#$'' in front of them. A {\em trending topic} is a term (one or more words) that in some moment appears in a huge amount of tweets of some place. These ``places'' and ``moments'' are known due to additional metadata contained in each tweet, which are related with geolocation, time and broadcaster account information, etc. Many times, trending topics are written as hashtags. 

\begin{table}[t]
  \begin{center}
  {\small
    \begin{tabular}{c|c|c}
          & user      & tweet    \\\hline
     user & \begin{tabular}{c}follows / is followed by\\mention\\replies to\\retweets to\end{tabular}
          & \begin{tabular}{c}posts\\retweets\\likes\\replies\end{tabular}\\\hline
    tweet & \begin{tabular}{c}posted by\\retweeted by\\liked by\\replied by\end{tabular}
          & \begin{tabular}{c}replies / is replied from\\retweets / is retweeted from\end{tabular}\\
    \end{tabular}
  \caption{Twitter relationships between users and tweets.\label{tab:TW-actions}}}
  \end{center}
\end{table}

In Twitter there are four types of public relationships: user-to-user, user-to-tweet, tweet-to-tweet and tweet-to-user. The valid actions for each kind of relationship are shown in Table~\ref{tab:TW-actions}. Note that both the user-to-tweet and the tweet-to-user relationships are symmetric. The {\em mention} action was already described before. In Twitter, when a user $A$ is following another user $B$, we say that $A$ is a {\em follower} of $B$, and $B$ is a {\em followee} of $A$. We can also {\em reply} a tweet of another user or, in other words, reply to another user through one of his tweets. Therefore, we can consider that the reply action belongs to the four kinds of relationships. Two very powerful actions are sharing, or {\em retweet}, the tweet of another user, and marking a tweet as {\em favorite} or {\em like}.\footnote{{\em Likes} are an evolution of the old {\em favorites} that were replaced at 2015. The differences between these two actions are very subtle, so for the purposes of the measures considered in this paper, we can consider them as analogous terms.} Retweets and likes also involve an implicit interaction between the user's action and the author of the original tweet. Since 2015, Twitter also provides the possibility of including text in retweets, so now we can both retweet and mention another user in the same tweet.

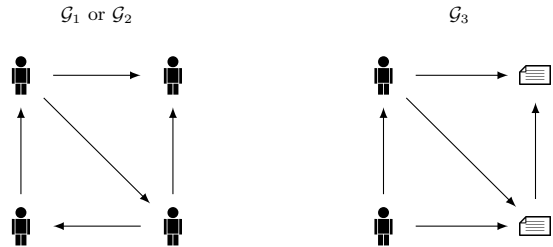
\begin{figure}[t]
\centering
\begin{minipage}[b]{0.25\linewidth}
\begin{center}
\begin{tikzpicture}[every node/.style={circle,scale=0.8}, node distance=7mm, >=latex]
\node[scale=1] at (1,2.8){$\cG_1$ or $\cG_2$};
\node[](a) at (0,2)[] {\Huge{\Gentsroom}};
\node[](b) at (2,2)[] {\Huge{\Gentsroom}};
\node[](c) at (0,0)[] {\Huge{\Gentsroom}};
\node[](d) at (2,0)[] {\Huge{\Gentsroom}};
\draw[->] (a) to node {}(b);
\draw[->] (a) to node {}(d);
\draw[->] (c) to node {}(a);
\draw[->] (d) to node {}(b);
\draw[->] (d) to node {}(c);
\end{tikzpicture}
\end{center}
\end{minipage}
\qquad
\begin{minipage}[b]{0.25\linewidth}
\begin{center}
\begin{tikzpicture}[every node/.style={circle,scale=0.8}, node distance=7mm, >=latex]
\node[scale=1] at (1,2.8){$\cG_3$};
\node[](a) at (0,2)[] {\Huge{\Gentsroom}};
\node[](b) at (2,2)[] {\Large{\PaperLandscape}};
\node[](c) at (0,0)[] {\Huge{\Gentsroom}};
\node[](d) at (2,0)[] {\Large{\PaperLandscape}};
\draw[->] (a) to node {}(b);
\draw[->] (a) to node {}(d);
\draw[->] (c) to node {}(a);
\draw[->] (d) to node {}(b);
\draw[->] (c) to node {}(d);
\end{tikzpicture}
\end{center}
\end{minipage}
\caption{Three different ways to represent Twitter as digraphs. The different kinds of relationships are described in Table~\ref{tab:TW-actions}. $\cG_1$ considers only follow-up relationships, while $\cG_2$ considers the more dynamic actions of mentions, replies and retweets. All the possible relationships can be represented by $\cG_3$.\label{fig:graphs}}
\end{figure}

As any social network, Twitter can be represented by a digraph $G=(V,E)$, where $V$ is the set of nodes and $E$ is the set of directed edges that represent how the nodes are related. The most traditional view is to consider that $V$ is the set of Twitter users, and the edges represent the follow-up relationship between users, so that an edge $(u,v)\in E$ means that user $u$ is a follower (or a followee, depending on the model) of user $v$. In such a case, the interactions among users like retweets, replies, mentions, etc., are part of the implicit dynamics of the network. A second possibility is to consider a digraph with the same set of nodes, but where the edges now represent a dynamic interaction between users, like for instance a retweet from one user to some tweet of another user. A more complete way of representation that may include the previous ones, requires a partition of the set of nodes into two sets $V_1$ and $V_2$, where $V_1$ is the set of users of the network and $V_2$ is the set of all tweets posted by the users. In this case, edges can represent the four kind of relationships~\cite{YTAK10}. The edges on the last two kinds of graphs are usually weighted, in order to represent the volume or frequency of the different interactions. In what follows, we denote these three kinds of graphs as $\cG_1$, $\cG_2$ and $\cG_3$, respectively. For simplicity, when a graph of kind $\cG_2$ only considers, for instance, edges for retweets and mentions, we say that it is a ``$\cG_2$ graph of retweets and mentions''. In Figure~\ref{fig:graphs} we illustrate these three kinds of graphs. Of course, there can be many other variations. For instance, Xie et al.~\cite{XPZL13} use a $\cG_2$ graph of reciprocal retweets, i.e., one that only has bidirectional edges that represent mutual retweets among users. Furthermore, a Twitter network can be represented by many other mathematical models, like multigraphs~\cite{JTB12}, Markov chains~\cite{SVH13}, edge-colored graphs as a particular kind of multilayer networks~\cite{ODA15}, among others.

\subsection{Twitter metrics}
\label{sec:metrics}

We consider a {\em metric} as a simple mathematical expression that helps us to provide basic information about the social network in the form of a numerical value. In turn, metrics can be combined to define a {\em (ranking) measure}, i.e., either a formula or an algorithm that provides a criterion to rank each user of a network. Of course, there exist measures which are more complex than a simple combination of metrics.

Pal and Counts~\cite{PC11} suggest a set of metrics that involve original tweets, replies, retweets, mentions and graph characteristics. After that, other researchers have continued developing this metric system~\cite{ZLCFC14}. We list these metrics in Table~\ref{tab-metrics}, by adding new metrics based on favorites or likes. First column of the table represents the metric identification, and the second column describes the computation of the metric for one particular Twitter user. Third column is the time complexity required to compute each metric, considering as input a Twitter database. This latter column is explained later in Section~\ref{sec:comcom}.

\begin{table}[t!]
\begin{center}
{\footnotesize
\begin{tabular}{c|l|l}
ID    & Metric description & Time complexity\\\hline\hline
$OT1$ & Number of original tweets (OTs) posted by the author. & $O(T)$\\\hline
$OT2$ & Number of URL links shared by their OTs. & $O(T)$\\\hline
$OT3$ & Number of hashtags included in their OTs. & $O(T)$\\\hline\hline
$RP1$ & Number of replies posted by the author. & $O(T)$\\\hline
$RP2$ & Number of OTs posted by the author and replied by other users. & $O(T)$\\\hline
$RP3$ & Number of users who have replied author's tweets. & $O(T\cdot k)$\\\hline\hline
$RT1$ & Number of retweets accomplished by the author. & $O(T)$\\\hline
$RT2$ & Number of OTs posted by the author and retweeted by other users. & $O(T)$\\\hline
$RT3$ & Number of users who have retweeted author's tweets. & $O(T\cdot k)$\\\hline\hline
$FT1$ & Number of tweets of other users marked as favorite (liked) by the author. & $O(T)$\\\hline
$FT2$ & Number of author's tweets marked as favorite (liked) by other users. & $O(T)$\\\hline
$FT3$ & Number of users that have marked author's tweets as favorite (likes). & $O(T\cdot k)$\\\hline\hline
$M1$  & Number of mentions to other users by the author. & $O(T)$\\\hline
$M2$  & Number of users mentioned by the author. & $O(T\cdot k)$\\\hline
$M3$  & Number of mentions to the author by other users. & $O(T)$\\\hline
$M4$  & Number of users mentioning the author. & $O(T\cdot k)$\\\hline\hline
$F1$  & Number of followers. & $O(1)$\\\hline
$F2$  & Number of topically active followers. & limited request\\\hline
$F3$  & Number of followees. & $O(1)$\\\hline
$F4$  & Number of topically active followees. & limited request\\\hline
$F5$  & Number of followers tweeting on topic after the author. & limited request\\\hline
$F6$  & Number of followees tweeting on topic before the author. & limited request\\
\end{tabular}
\caption{List of some valuable Twitter metrics with their respective time complexity. $T$ is the total number of tweets in the input, and $k$ the length of an auxiliary vector of unrepeated elements (see Section~\ref{sec:comcom}).}\label{tab-metrics}
}
\end{center}
\end{table}

Some of these metrics are relatively correlated. For instance, according to Ye and Wu~\cite{YW13}, if $RT2$ has a high value, then $RT3$ usually tends to be high; however, $M4$ tends to be low, even if $M3$ is high.

Note that $F2$ and $F4$ introduce topical analysis that involves analyzing the content of the tweets. This can be done by using machine learning techniques and natural language processing. In Section~\ref{sec:topical-inf} we survey the main influence measures related with topical analysis. The same metrics also refer to ``active'' users. As we shall see in Section~\ref{sec:inf-user}, ``activity'' is an ambiguous concept, like ``influence'' and ``popularity''. How often a user must tweet to be considered active? It depends on the context. As we mentioned, each tweet has its date of creation associated as metadata. Hence, we can construct a sequential timeline to see the order of the user's posts. Many other metrics of Table~\ref{tab-metrics} can also be restricted by topic and time~\cite{LLXY14}.

According to Chorley et al.~\cite{CCAW15}, metrics of retweets are the best quantitative indicators to prefer reading a tweet over another. This means that for a reader, a tweet being retweeted several times is more attractive than, for instance, a tweet with a lot of mentions. Despite of this, the most important indicators are qualitative, e.g., the friendship between the reader and the tweet's author. These considerations further hinder the definition of influential users.

\section{Twitter API}
\label{sec:API}
All the information required to compute these metrics is accessible through the Twitter API~\footnote{\url{https://dev.twitter.com/}}. An {\em API} ({\em Application Programming Interface}) is a set of functions, protocols and tools that are used to build an application, or to facilitate the communication with services. Twitter provides three kinds of API to developers (two of them are public) that together provide access to information from the social network. This information is obtained through different kind of requests, and it is useful for research, to search for historical or real-time information, and to develop unofficial clients.

\subsection{API REST}
\label{sec:API-rest}
{\em API REST} allows to read or write data on Twitter through simple HTTP primitives like {\em GET} and {\em POST}.\footnote{\url{https://dev.twitter.com/rest/public}} For instance, if a GET is performed to the ``\textsf{https://api.twitter.com/1.1/trends/place.json?id=1}'' resource, then the API will return 50 global trending topics. We show in Listing~\ref{RestAPI} an extract of the answer to the previous query. The answer is presented in JSON format and it consists of key-value pairs. Note that the value for a key can be a list containing more key-value pairs.

\begin{center}
{\scriptsize
\begin{minipage}{0.8\linewidth}
\begin{lstlisting}[language=json,firstnumber=1, caption= Extract of JSON response, label=RestAPI]
[
  {
    "as_of": "2012-08-24T23:25:43Z",
    "created_at": "2012-08-24T23:24:14Z",
    "locations": [
     {
       "name": "Worldwide",
       "woeid": 1
     }
    ],
    "trends": [
     {
       "tweet_volume": 3200,
       "events": null,
       "name": "#UserAccount",
       "promoted_content": null,
       "query": "%23UserAccount",
       "url": "http://twitter.com/search/?q=%23UserAccount"
     }, ...
\end{lstlisting}
\end{minipage}}
\end{center}

The main limitation for obtaining data through the API REST is the limited number of times we can do each kind of request. These numbers are controlled by tokens that are reduced for each kind of request, whenever a request of that kind is made. The amount of tokens to be performed may differs depending on whether we connect to the API with a user authentication or with an application authentication. Twitter provides to each developer a working window of 15 minutes, so if the tokens are depleted, we must wait to finish the current window, so that the number of tokens is restarted for another 15 minutes, and so on.

Twitter provides to the developers a list with the different types of requests and their associated tokens.\footnote{\url{https://dev.twitter.com/rest/public/rate-limits}}. For instance, the request \textsf{followers/ids} has 15 tokens, each of which returns a maximum of 5000 identifiers. Therefore, if we want to get the followers list of the Twitter account with the maximum number of followers to date (@KatyPerry, with 78,822,185 followers) we would need 15765 tokens, or 262.75 hours in the best case (Twitter does not ensure that the returned identifiers are always unique). After that, if we want more information about each follower, we must perform a GET to the resource ``\textsf{users/show}'', which returns complete information about a user. An extract of the data available through the resource ``\textsf{users/show}'' is shown in Listing~\ref{RestUser}.

\begin{center}
{\scriptsize
\begin{minipage}{0.8\linewidth}
\begin{lstlisting}[language=json,firstnumber=1, caption=Extract of JSON user response, label=RestUser]
  "favourites_count": 757,
  "follow_request_sent": false,
  "followers_count": 143916,
  "following": false,
  "friends_count": 1484,
  "geo_enabled": true,
  "id": 2244994945,
  "id_str": "2244994945",
  "is_translation_enabled": false,
  "is_translator": false,
  "lang": "en",
  "listed_count": 516,
  "location": "Internet",
   ...
  "statuses_count": 1279,
  "time_zone": "Pacific Time (US & Canada)",
  "url": "https://t.co/66w26cua1O",
  "utc_offset": -25200,
  "verified": true
  \end{lstlisting}
\end{minipage}}
\end{center}

Given the limitations of the API REST, it is possible that the representation of the network as a graph turns out to be prohibitive, because of the time needed for each request. For instance, let us consider to build a $\cG_1$ graph, assuming that each user has 5000 followers (so that for each user we only require one token to obtain all their identifiers). In this case, to obtain the second searching depth level, more than 83 hours are needed; for the third level, more than 17361 hours are needed, and so on, tending to an exponential curve. According to studies conducted by the consulting Beevolve and published in The Telegraph (2012)~\cite{Tel}, an average user has only 208 followers. Even so, to obtain the follow-up relationships between users in a third depth level we require more than 721 hours.

\subsection{Streaming API}

The {\em Streaming API}, also known as {\em Firehose}, provides developers access to the global flow of tweets and other events in real time (deletions, retweets, replies, etc.).\footnote{\url{https://dev.twitter.com/streaming/overview}} 
Unlike API REST, the Streaming API generates a persistent communication between the client and Twitter, which allows a continuous data stream. This API also provides various filters that allow to obtain data that only meet certain features (e.g., tweets sent from a certain geographical area, tweets that contain a specific bag of words, tweets containing hashtags, among others). Listing~\ref{StreamingAPI2} shows an extract of a Streaming API response.

\begin{center}
{\scriptsize
\begin{minipage}{0.8\linewidth}
\begin{lstlisting}[language=json,firstnumber=1, caption=Extract of Streaming API response, label=StreamingAPI2]
    "created_at": "Mon Nov 16 18:23:16 +0000 2015",
    "id": 666320647509921794,
    "id_str": "666320647509921794",
    "text": "RT @ZeikBaard: Het is de schuld van de wiet mensen. We zijn er. #dwdd",
    ...
    "in_reply_to_status_id": null,
    "in_reply_to_status_id_str": null,
    "in_reply_to_user_id": null,
    "in_reply_to_user_id_str": null,
    ...
    "user": {
        "id": 39451098,
        "id_str": "39451098",
        "name": "Pika",
        "screen_name": "TeshhCrack",
        "location": "Somewhere over the rainbow",
    ...
  \end{lstlisting}
\end{minipage}}
\end{center}

Years ago, Firehose allowed access to all Twitter data stream. However, since 2012 the new versions of the API began to restrict access to data~\cite{Sip12}. Nowadays, the Streaming API gives access to about 1\% of the total data.

Twitter originally sold tweets through three resellers: Gnip, Datasift and NTT Data. In 2014, Twitter acquired Gnip, and since then it sells tweets directly to its customers, exclusively through this service~\cite{Hof15}. Currently, the cost of tweets is very high, and there are profitable companies outside Twitter that are engaged in selling tweets. As an approach, from a price quote requested by the authors at the end of 2015, we observed that an annual subscription for downloading paid tweets can reach \$39,000 dollars.

These restrictions have changed the case studies of researches after 2012. The articles before that date, could easily obtain large databases of tweets that today are very expensive. On the one hand, due to the huge daily flow of tweets, this restriction of 1\% is still significant for case studies. On the other hand, these restrictions have led many researchers to turn their attention to Twitter analysis in real time. The latter, as discussed in Section~\ref{sec:influence}, has also motivated new ways to measure the influential actors within the network.

\subsection{Gnip}
Gnip\footnote{\url{https://gnip.com/}} is the business platform provided by Twitter to access their data. Unlike the two APIs aforementioned, this API is not public and prices vary according to the needs of each client.

This API allows to work with the total flow of events taking place in the social network (full details in real time). It ensures reliability data capture, through different fault tolerance mechanisms, such as resuming flow in case of a shutdown, the ability to use redundant connections across replicated servers, among others. Furthermore, it also offers an API oriented to requests, which is able to obtain historical information (through the use of complex rules), and also gets full access to the public tweets file.

\subsection{Anatomy of a tweet}
Despite the difference between Gnip and the public APIs, the latter also provide a large amount of data. The data accessible through the Streaming API can be classified into three groups. The first group is directly related to the tweets received. It provides information such as their unique identifiers, date of creation, sources (mobile application, web application, unofficial client, etc.), references to other social network features (if the tweet is a reply, retweet, mention, etc.), language, location, among others. The complete list of attributes for a tweet is the following:
\vspace{1ex}

{\small
\begin{tabular}{llll}
  \textsf{created\_at} & \textsf{id} & \textsf{id\_str} & \textsf{text}\\
  \textsf{source} & \textsf{truncated} & \textsf{in\_reply\_to\_status\_id} & \textsf{in\_reply\_to\_status\_id\_str}\\
  \textsf{in\_reply\_to\_user\_id} & \textsf{in\_reply\_to\_user\_id\_str} & \textsf{in\_reply\_to\_screen\_name} & \textsf{geo}\\
  \textsf{coordinates} & \textsf{place} & \textsf{contributors} & \textsf{is\_quote\_status} \\ 
  \textsf{retweet\_count} & \textsf{favorite\_count} & \textsf{favorited} &  \textsf{retweeted}\\
  \textsf{filter\_level} & \textsf{lang} & \textsf{timestamp\_ms} & \\[2ex]
\end{tabular}}

The second group refers to the user performing the action. The API returns a user object that provides his name, screen name, location, profile description, number of followers and followees, among others. The different attributes for a user are the following:
\vspace{1ex}

{\small
\begin{tabular}{llll}
 \textsf{id} & \textsf{id\_str} & \textsf{name}\\
 \textsf{screen\_name} & \textsf{location} & \textsf{url}\\
 \textsf{description} & \textsf{protected} & \textsf{verified}\\
 \textsf{followers\_count} & \textsf{friends\_count} & \textsf{listed\_count}\\
 \textsf{favourites\_count} & \textsf{statuses\_count} & \textsf{created\_at}\\
 \textsf{utc\_offset} & \textsf{ime\_zone} & \textsf{geo\_enabled}\\
 \textsf{lang} & \textsf{contributors\_enabled} & \textsf{is\_translator}\\
 \textsf{profile\_background\_color} & \textsf{profile\_background\_image\_url} & \textsf{profile\_background\_image\_url\_https}\\
 \textsf{profile\_background\_tile} & \textsf{profile\_link\_color} & \textsf{profile\_sidebar\_border\_color}\\
 \textsf{profile\_sidebar\_fill\_color} & \textsf{profile\_text\_color} & \textsf{profile\_use\_background\_image}\\ 
 \textsf{profile\_image\_url} & \textsf{profile\_image\_url\_https} & 
 \textsf{profile\_banner\_url}\\
 \textsf{default\_profile} & \textsf{default\_profile\_image} & \textsf{following}\\
 \textsf{follow\_request\_sent} & \textsf{notifications} & \\[2ex]
\end{tabular}}

The third group refers to the categorization of entities and considers the tweet's content:
\vspace{1ex}

{\small
\begin{tabular}{lllll}
 \textsf{hashtags} & \textsf{symbols} & \textsf{urls} & \textsf{user\_mentions} & \textsf{media}\\[2ex]
\end{tabular}}

This anatomic description is based on the Streaming API. However, the information obtained from the API REST is quite similar. The main difference is the ability to get more information by performing historical searches according to some criteria and filters.

\section{What is an influential user?}
\label{sec:inf-user}

The problem of measuring the influence of a user in a social network is, in the first place, a conceptual problem. There is no agreement on what is meant by an influential user. Therefore, new influence measures are constantly emerging, each of which offers different measurement criteria.

The wide variety of influence criteria also involves the definition of new kinds of users who are closely related to influential users. Sometimes, influential users are also called {\em opinion leaders}, {\em innovators}~\cite{CXZW13}, {\em prestigious}~\cite{Gay13} or {\em authoritative actors}~\cite{BR15}. Occasionally, they have been associated with {\em topical experts} for specific domains~\cite{LLXY14,SZJ14}. In addition, there are other user classifications related to spread of influence. For example, some researchers distinguish between {\em opinion leaders}, {\em influencers} and {\em discussers} by type of activity and impact~\cite{JTB12}. We can also distinguish between {\em inventors} (users who start a new topic) and {\em spreaders} (users who are responsible for disseminating that topic)~\cite{LPLSLX14}. Another classification considers {\em disseminators} (users who spread their influence and prevent the formation of structural holes), {\em engagers} (users who facilitate relationships with third parties) and {\em leaders} (top disseminator-engagers)~\cite{FDS16}. We can also recognize {\em idea starters} (users with many followers) and {\em connecters} (users connecting starters)~\cite{SN13}, {\em amplifiers}, {\em curators}, {\em commentators} and {\em viewers}~\cite{TCHB12}. Influential users can also be classified according to their content and authority~\cite{XNT14}.

Other relevant users are the {\em celebrities}~\cite{SST13}, who meet somewhat different criteria from those of influencers. Users can be classified according to their popularity into {\em broadcasters} or {\em passive users}~\cite{RA14} (many followers and few followees), {\em acquaintances} (similar amount of followers and followees) and {\em miscreants} or {\em evangelists} (few followers and many followees, like spammers and bots)~\cite{KGA08}. For some researchers, the previous notion of evangelists corresponds to {\em mass media} users, keeping the name {\em evangelists} to well connected opinion leaders, in contrast to normal users or {\em grassroots}~\cite{CBHG12}. Besides the above, some researchers distinguish between {\em popular}, {\em influential}, {\em listener}, {\em star} and {\em highly-read} users, considering not only the existing metrics but also the tweets' content, as well as both their cognitive and psychological behaviour~\cite{QECC11}.

On the negative side, {\em meat-puppets} are a negative evolution of {\em sock puppets} in OSNs~\cite{WAHRC13}. Furthermore, {\em social capitalists} are users who only seek followers and retweets, but differ from spammers and bots~\cite{MSOB13} (see also Dugu\'e and Perez~\cite{DP14}; Danisch et al.~\cite{DDP14}). The spammer detection is a separate issue that has also received much attention on Twitter. Although this is not the main focus of the article, it is interesting to mention that some metrics can be used to recognize spammers on Twitter. For instance, if a user is retweeting or marking as favorite his own tweets, this could be considered an act of spam. Moreover, Pal and Counts~\cite{PC11} proposed an additional metric (not included in Section~\ref{sec:metrics}, because it is more complex than the others) called the self-similarity score, that computes the similarity between the recent and previous tweets of a users. If the self-similarity is high, the user can be a spammer; if it is low, the user speaks about many topics and uses a large vocabulary. According to Gayo-Avello~\cite{Gay13}, a good measure should place the most relevant users at the top of the ranking, without caring about their exact location in the ranking. At the same time, that measure should punish the spammers, placing them in lower positions.

It is important to note that an influential user is not necessarily one who writes influential tweets~\cite{RS15}. According to Francalanci and Hussain~\cite{FH15}, while influential users are typically associated with hub nodes, the influence can be spread through multi-layered peripheral node clusters. The tweet's influence can be measured in terms of {\em extension} (number of users ``affected'' by the message), {\em intensity} (cognitive, emotional and conductual impact)~\cite{MB14}, content, sender's popularity, etc. In a sociological sense, some researchers argue that opinion leaders in Twitter are more associated with people with high self-esteem~\cite{Hwa15}.

Li et al.~\cite{LPLSLX14} define a framework to observe the behavior of the influence between users. This observation considers three dimensions: the degree of similarity of the initiated topics, the latency between one tweet and the next on the same topic, and the tendency of a user to create new topics or spread already existing topics through the network.

The present list of user types is not exhaustive, but it allows us to realize that there is no standard definition to measure influence. Moreover, it seems reasonable to ensure that some of these types of users, such as celebrities and social capitalists, have no direct relation with influence criteria. In what follows, we propose to split the existing measures into three different kinds of criteria: {\em activity measures}, {\em popularity measures} and {\em influence measures}. Thus, for instance, the $RT2$ metric seems to be much more proper to measure influence than $F2$~\cite{WSHC11}, while $F1$ and $F3$ are useful metrics to measure popularity~\cite{SVH13}. This classification is valid because there is no direct correlation between active, popular and influential actors~\cite{RGAH11}.

\section{Activity measures}
\label{sec:activity}

We say that users are {\em active} when their participation in the social network is constant and frequent in a period of time, regardless of the attention they receive for their participation. Note that there may be very active Twitter's readers, who however cannot be observable by any metric, because they do not leave any trace on the network. Therefore, by ``participation'' we mean to do actions that can be measured, like doing tweets, retweets, mentions, replies, etc. In this sense, Yin and Zhang~\cite{YZ12} define the activity of a user as the probability of a user seeing a tweet. We cannot determine when a user has seen a tweet, but if a user retweets a tweet, we can assume that the tweet has been read. Therefore, more active users are more capable to see new tweets, and thus interact with them.

Perhaps the simplest activity measure is the {\em TweetRank}~\cite{NTC10}, which is just a metric that counts the number of tweets of the user. Slightly more sophisticated is the {\em Tweet count score}~\cite{NRXT13}, that counts the number of original tweets plus the number of retweets. Following this approach, a reasonable activity measure could be the sum of the visible actions of each user. Thus, for every user $i$ we define the {\em General Activity} as follows:
$$\mbox{{\em General Activity}}(i)=OT1+RP1+RT1+FT1$$
We assume that metric $M1$ is contained in the other metrics. We can normalize this measure by dividing the value by the total number of the considered tweets. Furthermore, we can restrict our attention to the tweets related with some specific topic. This is what the {\em topical signal} ($TS$)~\cite{PC11} does, in spite of it omits the likes:
$$TS(i)=\frac{(OT1+RP1+RT1)\Big|_{\mbox{\footnotesize{specific topic}}}}{\#\mbox{tweets}}$$
The same authors define a {\em signal strength} ($SS$)~\cite{PC11}, that indicates how strong is the author's topical signal:
$$SS(i)=\frac{OT1}{OT1+RT1}$$
This expression measures the originality of the author's tweets, in such a way that a greater authorship means values closer to $1$.

There are also measures that, although were defined as influence measures, under our classification they fit better as activity measures. Such is the case of the sum of {\em effective readers} for all user tweets~\cite{LKPM10}, where an effective reader of a tweet is a follower who still has not tweeted on any trending topic when the user sent the tweet. In some way, it measures the speed of a user to tweet about new topics. Note that despite being one of the oldest measures (it was published in 2010), it already takes into account the timeline of tweets. Note also that, unlike previous measures mentioned above, it cannot be calculated immediately, so it is more suitable for offline analysis. This is also the case of all the remaining measures in this section. We address the computational complexity aspects of measures in Section~\ref{sec:comcom}.

The {\em ActivityScore}~\cite{YLLH13} also considers the timeline. This measure counts the number of followers, followees and tweets on a $\cG_3$ graph for each user during a period of time. The {\em DiscussRank}~\cite{JTB12} determines how active a user is, in the sense of initiating conversations around a topic. This algorithm differs from the others in that it is based on multigraphs (i.e., graphs with multiple edges between two nodes), whose nodes are the users, and the edges are based on followers, retweets and mentions among users. An additional topical-sensitive measure is the {\em Competency}~\cite{AKRO15}, which ranks the actors according to his ability to post relevant tweets according to hot topics. The topical detection is given by a Latent Dirichlet Allocation (LDA) algorithm, and the topical relevance of each tweet is determined by a Divergence From Randomness retrieval model. One problem of Twitter measures based on LDA is that the traditional LDA algorithms are used for larger texts than tweets, so it is necessary to use specific variations~\cite{ZJWHL11}.

Finally, the {\em IP Influence}~\cite{RGAH11} not only measures the users' influence (see Section~\ref{sec:influence}), but also their passivity. The {\em passivity} of a user is defined as the difficulty for the user to be influenced by another in some period of time. It is measured by considering metrics of retweets, followers and followees. Most social networking users are not active, but passive, in the sense that they do not cooperate with the spread of content over the network. According to the authors, most users with very high passivity tend to be spammers or bots. This notion of passivity is also used in the {\em influence graphs}, where it is displayed on the labels of the nodes~\cite{MRS15}. The IP Influence was criticized by Gayo-Avello et al.~\cite{GBFFG11} because it does not consider the followers, and it was also disproved that it really correlates with the number of clicks on URLs.

\section{Popularity measures}
\label{sec:popularity}

Briefly speaking, we can say that a user is {\em popular} when he is recognized by many other users on the network. An example of a popular user is a celebrity, who does not necessarily have an active and influential account. For instance, the Clint Eastwood's account (@Eastwood$\_$) has more than 60,000 followers, but no followees, no tweets.

The simpler popularity measures just count the follow-up relationships between users. The {\em FollowerRank}~\cite{NTC10}, also known as {\em Structural Advantage}~\cite{CS12}, is the normalized version of the traditional {\em in-degree} measure~\cite{HW11,JW13} for social networks in general:
$$\mbox{{\em FollowerRank}}(i)=\frac{F1}{F1+F3}$$
There are variations of this measure, as the {\em Twitter Follower-Followee ratio (TFF)}~\cite{BCMGA12} that corresponds to $F1/F3$. The disadvantage of these measures is that $F1$ and $F3$ metrics may differ a lot, and the number of followers for some Twitter users is too high compared to the rest. To mitigate these differences, the {\em Popularity}~\cite{AKRO15} was defined as follows:
$$\mbox{{\em Popularity}}(i)=1-e^{-\lambda\cdot F1}$$
where $\lambda$ is a constant that by default can be equals $1$. As both metrics $F1$ and $F3$ can be obtained directly from the API REST, all the previous popularity measures can be computed in real time. However, note that with few changes, these measures lose their efficiency and become on measures restricted to offline environments. This is the case of the {\em Followers to followee ratio with paradoxical discounted reciprocity}~\cite{Gay13}, that introduces an additional metric, namely the number of reciprocal actors of a user, i.e., the number of followers who are also followees:
$$\mbox{{\em Paradoxical discounted}}(i)= \left\{
     \begin{array}{cl}
       F1/F3 & \mbox{if }F1>F3\\
       \frac{F1-\mbox{reciprocal}(i)}{F3-\mbox{reciprocal}(i)} & \mbox{otherwise}\\
     \end{array}
 \right.$$
The purpose of this measure is to punish the spammers (users with many followees and few followers). However, computing the reciprocal$(i)$ metric is not as simple as computing $F1$ and $F3$. Indeed, it considerably increases the computational costs, as we mention in Section~\ref{sec:API-rest}. A simple topic-sensitive popularity measure, also based on complex metrics (see Table~\ref{tab-metrics}), is the {\em Network Score} $(NS)$~\cite{PC11}, which is based on the active non-reciprocal followers of the user:
$$NS(i)=\log(F2+1)-\log(F4+1)$$

All popularity measures seen so far consider metrics based on follow-up relationships. However, popularity measures may also consider other kind of metrics. A good example is given by three measures defined by Srinivasan et al.~\cite{SST13}. Each one of these measures is based on the previous one, in addition to other metrics. Despite the variety of metrics used for the first two, only the third one becomes more complex, because of the API limitations discussed above. The first one is the {\em Acquaintance Score} $A(i)$ that measures how well-known user $i$ is. Let $n$ be the number of considered user accounts, it is defined as:
$$A(i)=\frac{F1+M4+RP3+RT3}{n}$$
The second measure is the {\em Acquaintance-Affinity Score} $AA(j)$ that measures how dear user $j$ is, by considering how well known are those who want him:
$$AA(j) = \sum_{i\in E_{RP}}A(i)\cdot\frac{\#\mbox{replies of }i\mbox{ to }j}{\#\mbox{replies of }i}
+ \sum_{i\in E_M}A(i)\cdot\frac{\#\mbox{mentions of }i\mbox{ to }j}{\#\mbox{mentions of }i}
+ \sum_{i\in E_{RT}}A(i)\cdot\frac{\#\mbox{retweets of }i\mbox{ to }j}{\#\mbox{retweets of }i}$$
where $E_{RP}$, $E_M$ and $E_{RT}$ are the set of users who reply, mention and retweet the tweets of $j$, respectively.
Finally, the {\em Acquaintance-Affinity-Identification Score} $AAI(j)$ measures how identifiable user $j$ is, by considering how dear are those who identify him:
$$AAI(j) = \sum_{i\in F_r}\frac{AA(i)}{\#\mbox{followees of }i}$$
where $F_r$ is the set of followers of $j$. The AAI Score is well correlated with the $F1$ metric, and was used to identify celebrities in the ``real world'', i.e., outside the Twitter network~\cite{SST14}. After that, the same authors defined the {\em Action-Reaction (AR)}~\cite{SST14} to identify celebrities within the Twitter network. It is a combination of two measures: the {\em Action score} that measures the loyalty of the user's ``fans'', and the {\em Reaction score} that measures the attention aroused by the celebrity for his actions. This measure uses conditional probability variables based on replies, mentions and retweets, and successfully avoids the complexity limitations of the AAI Score.

We finish this section with a measure based on the {\em PageRank}, a widely used algorithm to define centrality measures, and specially influence measures, as we will see in the following section. The {\em Starrank}~\cite{KC10} uses PageRank to make a daily dynamic analysis in a $\cG_2$ graph of mentions. It considers metrics such as the acceleration of mentions over time. This popularity measure is interesting, because it does not consider explicitly the follow-up relationships, but the times the user is mentioned by other Twitter accounts.

\section{Influence measures}
\label{sec:influence}

In this section we survey the different influence measures for the Twitter network, which is the main purpose of this paper. We say that a user is {\em influential} whether his actions in the network are capable to affect the actions of many other users in the network. Since we are focused on a microblog service, we can also understand that influential users are the most able to spread information within the network~\cite{MM15}. The influential users tend to be active~\cite{CHBG10,KC10}, but only a few active users are influential. Furthermore, it is important to understand that influential users in a social network are not neccesarily influential in the real life~\cite{CDL15,CLD15}.

According to Quercia et al.~\cite{QECC11}, there are two broad paradigms of social influence. The first point of view believes that the influence is often massively exerted by a small number of very persuasive or connected users. The second point of view believes that many users can be influential accidentally, depending on many unpredictable factors. Despite the second paradigm, the authors argue that for social networks like Twitter it is possible to develop several influence criteria that can be quantified. Morone and Makse~\cite{MM15} are inclined to the first of these two points of view, although they suggest that for complex networks, the influential actors also respond to unpredictable criteria, and usually are not those who are better connected. Borge-Holthoefer et al.~\cite{BBGM13} also agree with the first point of view, but in addition, they support the possibility that the influence accumulated in smaller subnetworks, formed by less influential users, could explode with a cascading behaviour, affecting the whole network.

\subsection{Traditional measures used on Twitter}

In Section~\ref{sec:metrics} we mentioned several metrics that can be determined from the particular dynamics of the Twitter network. The Twitter influence measures in literature usually consider metrics related to retweets, mentions and, to a lesser extent, followers~\cite{CHBG10}. However, some researchers have used traditional centrality measures that are not based on Twitter metrics at all, but only on the passive topology of the $\cG_1$ graphs. That is the case of {\em Closeness} $(C_c)$ and {\em Betweenness} $(C_B)$~\cite{HW11,JW13}. Closeness centrality is based on the length of the shortest paths from a node $i$ to everyone else. It measures the visibility or accessibility of each node with respect to the entire network. Let $D$ be the distance matrix of a network with $n$ nodes, it is defined as follows~\cite{LM07}:
$$C_c(i)=\frac{n-1}{\sum_{i\neq j}(D)_{ij}}$$
If there is no path from $i$ to $j$, then we assume that $(D)_{ij}=n$.
Betweenness centrality considers for each node $i$ all the shortest paths that should pass through $i$ to connect all the other nodes in the network. It measures the ability of each node to facilitate communication within the network. Let $b_{jk}$ be the number of shortest paths from node $j$ to node $k$, and $b_{jik}$ the number of those shortest paths that pass through node $i$, then~\cite{LM07}:
$$C_B(i)=\frac{1}{(n-1)(n-2)}\sum_{j\neq k}\frac{b_{jik}}{b_{jk}}$$

A relatively recent measure is the {\em Hirsch Index} or {\em H-index}~\cite{Hir10}, originally defined to measure the productivity of a person in the scientific community, according to the citations of his articles. In the context of Twitter, it can be defined as the maximum value $h$ such that $h$ tweets of the user have been replied, retweeted, or liked, at least $h$ times. A simplified version for Twitter was proposed by Romero et al.~\cite{RGAH11}, where instead of replies, retweets and likes, they only consider retweets of tweets that contain URLs. Razis and Anagnostopoulos~\cite{RA14} consider two additional versions: one based on retweets, and the other on favorites or likes. In order to do daily analysis reports, the authors take, for both measures, only the last one hundred daily tweets from each user.

Several influence measures that we consider in the next section are based on {\em PageRank}~\cite{PBMW99}, a well known algorithm used to measure both the relevance and presence of websites on the Internet. This algorithm considers the hub users as important nodes within the network. PageRank is a variation of another traditional centrality measure known as {\em eigenvector}~\cite{LM03}, that favors users who are well connected with other well connected users within the network~\cite{ST11}. An alternative algorithm to PageRank is {\em HITS}~\cite{Kle99}. While the former is executed at indexing time, the latter is executed at query time. Moreover, HITS provides two closely related scores: an authority score and a hub score. Both algorithms have been used repeatedly in the context of Twitter~\cite{Tun09,KLPM10,CLCW11,KF11,JTB12}. PageRank has been applied on $\cG_2$ graphs of retweets and replies, for dynamic analysis over different periods of time~\cite{ZSL11}. Both PageRank and HITS have been applied on $\cG_2$ graphs of retweets, replies and mentions for networks with sparse followers~\cite{PAJVS12}. In this latter case, authors also use, as an additional method, a decision tree classification model that classifies the nodes after a supervised training. Another somewhat less known algorithm is {\em NodeRanking} that, unlike PageRank, it can be used on graphs with weighted edges~\cite{PSD02}. For graphs without weights, NodeRanking and PageRank are almost equal~\cite{Gay13}. As PageRank, HITS and NodeRanking are recursive algorithms, and the Twitter network is so dynamic, measures based on these algorithms do not allow real-time analysis~\cite{CS12}. Despite of this, the smart idea behind the PageRank algorithm has been applied on several Twitter influence measures for offline analysis. The HITS algorithm, instead, is not advisable to be used on OSNs, since it gives too much power to spammers~\cite{Gay13}.

Besides the traditional centrality measures, some authors have been inspired on other disciplines different from social network analysis and computer science. Thus, Gayo-Avello et al.~\cite{GBFFG11} proposed an influence measure for Twitter based on the variable-mass system of physics. Here, the user score is represented by the velocity of an object over a period of time:
$$v_t = v_{t-1}+\frac{F_a}{m}-c$$
where time $t$ is taken per each $1$ hour (although in practice it can be calculated almost in real time), the force applied $F_a$ is the number of mentions made to the user during the last hour, the mass $m$ is the number of followers of the user (this value can be smoothed with a logarithm), and $c$ is a constant that can be zero. Negative velocities are also considered zero. The authors found a positive correlation with the number of clicks on URLs, analogous to IP Influence, PageRank and TunkRank (a direct adaptation of PageRank that we consider later). This correlation increases when the number of followers is adjusted.

We finish this section by mentioning three traditional methods used to deal with complex networks. The first method is the k-shell decomposition algorithm, that has been applied to identify influential actors in epidemic dynamics~\cite{KGHLMSM10}. Although this algorithm has been used in Twitter  directly~\cite{PMAZM14}, Borge-Holthoefer and Moreno~\cite{BM12} suggest that it does not work to measure influential spreaders in rumor dynamics. Through numerical simulations, they notice that well connected users may already know a recent news but are not willing to spread it anymore. A variation of this algorithm to identify influential users in a $\cG_1$ graph was proposed by Brown and Feng~\cite{BF11}. The second method is the particle swarm optimization (PSO) algorithm, that has been used with metrics of followers and retweets~\cite{ZZWLG13}. Finally, the {\em F-measure}~\cite{Rij79}, used in statistical analysis for binary classification, can be applied to study the power of retweets, by comparing the follow-up relationships and different centrality measures, such as betweenness, closeness, eigenvector and PageRank~\cite{WQP12}.

\subsection{Measures based on Twitter metrics and PageRank}
\label{sec:metric-pagerank}

Two simple measures based on the metrics of Table~\ref{tab-metrics} are {\em Retweet Impact} $(RI)$ and {\em Mention Impact} $(MI)$~\cite{PC11}. The first one estimates the impact of the user tweets, in terms of the retweeted tweets:
$$RI(i)=RT2\cdot\log(RT3)$$
The logarithms moderate the impact of overly enthusiastic users who retweet the same content many times. This measure could be easily generalized, by considering the same expression for the $FT2$ and $FT3$ pair of metrics. The second measure estimates the impact of the user tweets, in terms of the mentions received by other users:
$$MI(i)=M3\cdot\log(M4)-M1\cdot\log(M2)$$
The subtraction ensures that the mentions are given by their own merit, and not because the user has been mentioning other users. A third measure based on simple metrics is the {\em Social Networking Potential (SNP)}~\cite{AK11}, defined as follows:
$$\mbox{{\em SNP}}(i)=\frac{Ir(i)+RMr(i)}{2}$$
where the {\em Interactor Ratio}, $Ir(i)$, and the {\em Retweet and Mention Ratio}, $RMr(i)$, are defined as:
$$Ir(i)=\frac{RT3+M4}{F1} \hspace{.5cm} \mbox{ and } \hspace{.5cm} RMr(i)=\frac{\#\mbox{tweets of } i \mbox{ retweeted} \, \, + \, \, \#\mbox{tweets of } i \mbox{ replied}}{\#\mbox{tweets of } i}$$
For each user, $Ir(i)$ measures how many different users interact with that user, while $RMr(i)$ measures how many of his tweets imply a reaction from the audience.

Note that the SNP measure considers all kind of actions on Twitter, except the favorites or likes. The same occurs for a measure based on {\em content} and {\em conversation}, informally defined by Hatcher et al.~\cite{HBV11}. The content criterion considers the number of published tweets and follow-up relationships, and it has the $25\%$ of the total importance. The conversation criterion considers the number of replies, as well as the number of followers related with the user through mentions and retweets, and it has $75\%$ of the total importance of the measure.

Besides the previous measures, which are based directly on simple metrics, some authors have studied user influence in terms of their ability to post websites that are propagated through the network. This can be implemented using PageRank~\cite{BHMW11} and HITS. The {\em IP Influence}~\cite{RGAH11} (also mentioned in Section~\ref{sec:activity}) uses HITS in this context, in order to analyze the influence and passivity (based on retweets) of users over time.

The remainder of this section is devoted to several influence measures based on PageRank. The first direct adaptation of this algorithm into the context of Twitter was {\em TunkRank}~\cite{Tun09}:
$$\mbox{{\em TunkRank}}(i)=\sum_{j\in\mbox{\small{followers}}(i)}\frac{1+p\cdot\mbox{TunkRank}(j)}{\#\mbox{followees of} j}$$
where, analogously to PageRank, $0\leq p\leq 1$ is the probability that a tweet is retweeted. This probability is assumed to be equal for all users. In the literature, the authors normally use $p=0.5$, but in fact this value should vary from case to case. For instance, for experiments considered by Gayo-Avello~\cite{Gay13}, the ideal value was $p=0.287$. A variation of TunkRank is {\em UserRank}~\cite{MS12}, defined to measure the influence of a user according to the relevance of his tweets:
$$\mbox{{\em UserRank}}(i)=\sum_{j\in\mbox{\small{followers}}(i)}\frac{1+\frac{\#\mbox{\small{followers of }} i}{\#\mbox{\small{tweets of }} i}\cdot\mbox{UserRank}(j)}{\#\mbox{\small{followers of }} j}$$
Note that these measures changes the denominator $\mbox{followees}(j)$ by $\mbox{followers}(j)$.
As PageRank, TunkRank is correlated to the number of clicks on URLs~\cite{GBFFG11}. To penalize the ranking of spammers, Gayo-Avello~\cite{Gay13} proposed, for each user $i$, to multiply PageRank$(i)$ by Paradoxical$\_$discounted$(i)$ (see Section~\ref{sec:popularity}). This could also be done with the TunkRank. Indeed, as TunkRank is an adaptation for Twitter, this measure tends to penalize spammers more than PageRank~\cite{Gay13}.

Besides spammers, there are users using multiple false accounts. These accounts can be used to interact among themselves, and thus to improve their influence scores. The {\em TrueTop}~\cite{ZZSZZ15} measure aims to be resilient to sybil attacks. It applies weighted eigenvector centrality (recall that PageRank is an evolution of eigenvector centrality) in a $\cG_2$ graph of retweets, replies and mentions, where the edge weights represent the number of such user interactions. The algorithm allows to split the graph in a region with sybil users and another region without them, so that the users in the second region receive a higher score in the ranking.

The following measures strongly depend on follow-up relationships among users, so that the acquisition of necessary data is restricted to the limitations of Twitter API. The {\em diversity-dependent influence score (DIS)}~\cite{HLLC13} considers a variation of PageRank on a $\cG_2$ graph of retweets and following. To favor a deeper influence spread, it gives more power to those users that are capable to influence (according to their follow-up relationships) the less dense users, i.e., those with a ``high diversity''. The authors also proposed a variation in the definition of ``diversity'', allowing a dynamic computation of the measure. The {\em Influence Rank (IR)}~\cite{HW11} combines follow-up relationships, mentions, favorites and retweets, to identify opinion leaders who are capable of influencing other influential users. Once the graph is constructed, it can be implemented as a polynomial approximation algorithm based on PageRank. At last, Li et al.~\cite{LCCJ13} proposed that a user $A$ can exert more influence over another user $B$ if $A$ writes tweets more related with those that $B$ often writes. With this idea, the authors use a variation of PageRank based on a similarity factor between published tweets, on a $\cG_2$ graph of following, retweets, mentions and replies. This latter idea brings new considerations to state the influence of users. However, this by itself does not seem to be enough. Due to the spontaneous phenomenon of viral content, it makes sense to think that a user can be influential without reading constantly what others write. For some authors, the user profile and the style of the tweets can be an additional issue to be considered. In this line, Ram\'irez-de-la-Rosa et al.~\cite{RVJS14} proposed 23 features of user profile (presence of hashtags, URLs, self-mentions, number of followers and tweets, etc.) and 9 features of tweet's style (extension, frequency, quality, number of retweets, etc.) that for them are involved in influence capacity.

We continue with two measures called {\em InfRank} and {\em LeadRank}~\cite{JTB12} InfRank is a variation of PageRank that measures the user influence in terms of his ability to spread information and to be retweeted by other influential users. The algorithm operates on a multigraph whose nodes are the users and the edges are based on retweet relationships. LeadRank measures the leadership of a user, in terms of his ability to stimulate retweets and mentions from other users on the network, especially other leaders. This algorithm operates on a similar multigraph, but here the edges can represent retweets and mentions. Although these measures do not consider the follow-up relationships, they are also restricted by the Twitter API limitations.

Until now we have not seen any measure based on PageRank that avoids the limitations of the Twitter API. Two measures that bypass these limitations are {\em SpreadRank}~\cite{DJZHHZ13} and {\em ProfileRank}~\cite{SGMZ13}. The first one measures the spreadability of users on a $\cG_2$ graph of retweets, whose edges are weighted by the proportion of retweets with respect to the total tweets of the retweeted user. The authors suggest that spreading influence is greater the faster their tweets are retweeted. They also consider information cascades, which can be described as a tree that grows over time, and whose nodes are the retweets of an original post. The closer the user is to the root node (earlier in time), the greater his influence spread. Therefore, the measure takes into account both the time interval between retweets and the location of users in the information cascade. In turn, the ProfileRank considers the influencers as users that can generate relevant content to other users. Inspired on PageRank, it is computed by random walks on a bipartite $\cG_3$ graph whose edges represent generation and consumption of content by users over time.

Some ideas mentioned above have also been used by other authors. Valiati et al.~\cite{VSGM13} adapt PageRank on bipartite $\cG_3$ graphs of retweets. Ding et al.~\cite{DJZH13} define {\em MultiRank}, a more complex measure inspired on PageRank that considers random walks on $\cG_2$ graphs of retweets, replies, and two additional relationships, namely ``reintroduce'' and ``read''. ``Reintroduce'' refers to post tweets which are similar to others previously posted by another user, but without acknowledging that user as the source. ``Read'' refers to the probability to read tweets published by other users, according to the appearance of such tweets in their timelines. In addition, the authors characterize the spammers in order to rank them below the common Twitter accounts.

Finally, there are influence measures that extend the PageRank algorithm. {\em TURank}~\cite{YTAK10} is based on {\em ObjectRank}, which is indeed an extension of PageRank. The algorithm uses a $\cG_3$ graph of following, tweeting and retweets. From a similar graph that also allows replies, Liu et al.~\cite{LWH13} add a ``time-effectiveness attenuation coefficient'' (TAC) that returns the time on which tweets are published, so that the earliest tweets gradually lose relevance in the measure. Again, these measures are restricted by Twitter API limitations.

\subsection{Topical influential users}
\label{sec:topical-inf}

Several authors deal with the problem of identifying influential users regarding some specific topic. For these cases we need to analyze the content of the tweets. Traditional LDA algorithms and topical modeling tools are often more helpful in larger texts than tweets, so for this context it is necessary to use specific variations of the algorithms~\cite{ZJWHL11}. These computational techniques are not the purpose of this survey, so they will not be detailed here. Somehow, topical-sensitive measures are a kind of local measures, not in a topological sense but in a semantic level. According to Kardara et al.~\cite{KPPTV15}, these measures are more effective and functional than global measures, i.e., those that are not restricted to a topical domain.

These measures are usually used in offline analysis. A volume of tweets is collected by streaming over a certain period of time. During or after obtaining a database, tweets can be filtered to leave only those related to a specific topic. By using this method, Schenk and Sicker~\cite{SS11} analyzed the variation of followers who posted tweets about a specific natural disaster. For the authors, the most influential users are those with a greater positive variation.

At least two traditional centrality measures can be easily adapted as topical-sensitive measures for Twitter. The {\em alpha centrality}~\cite{BL01}, based on eigenvector and the time as an additional parameter, has been adapted to consider retweets~\cite{OGPJ13}. In turn, the {\em T-index}~\cite{KRGVBD13} is a topical variation of $H$-index and also considers retweets.

We found only one measure defined as a combination of metrics of Table~\ref{tab-metrics}. The {\em Information Diffusion}~\cite{PC11} estimates the possible influence of the user's tweets, among his followers who are non-followees:
$$ID(i)=\log(F5+1)-\log(F6+1)$$
The ``$+1$'' in the logarithms avoids divisions by zero, in case we want to normalize the measure. This measure only considers follow-up relationships, but it is independent of the number of followers and followees on the network. Besides $F5$ and $F6$ metrics, $F2$ and $F4$ are also topical-sensitive. Therefore, every measure based on some of these metrics is affected by Twitter API limitations.

There are other measures affected by these limitations. The {\em Topic-Specific Author Ranking}~\cite{KF11} considers the quality of the tweets posted on a given topic. The quality of each tweet is measured in terms of its interest in the network, by using metrics of mentions and retweets. The algorithm is applied on a $\cG_3$ graph of following, tweets and retweets (similar to that used in TURank). The implementation uses the MapReduce framework, and it is compared with some rankings based on PageRank and HITS. On the other hand, Sun et al.~\cite{SZL13} define six measures based on effective audience for each user. The effective audience is quantified in terms of follow-up relationships, retweets and replies. Two of these measures consider a topical analysis, and another provides a dynamic value that depends on time. The first one has a high correlation with PageRank on $\cG_1$ and $\cG_2$ graphs of retweets, with the $F1$ metric, and especially with the $RT1$ metric.

Instead of consider the full text of the tweets, some measures restrict their attention to the hashtags. Such is the case of {\em TRank}~\cite{MF15}, that ranks users in three dimensions by considering follow-up relationships, retweets and favorites. They also consider human factors, such as the fact that a user with many active followees may simply not be able to read all the tweets they publish. Other examples, without API limitations, are {\em RetweetRank} and {\em MentionRank}~\cite{XNT14}, restricted to hashtags of news events. Under the assumption that interaction with influential users makes you more influential, they extend PageRank on $\cG_2$ graphs of retweets and mentions, respectively.

As expected, PageRank can also be applied in this context. Cataldi and Aufaure~\cite{CA14} proposed a direct topic-oriented adaptation of PageRank on $\cG_2$ graphs of retweets, that requires the number of users of the network.

The {\em Topic-sensitive PageRank (TSPR)}~\cite{Hav03} was created as an extension of PageRank for the original purpose of ranking websites. The {\em TwitterRank}~\cite{WLJH10} uses a similar idea as TSPR for the context of Twitter. It is applied on $\cG_1$ graphs and considers criteria of similarity between users, restricted to the topics of their tweets. The TwitterRank algorithm is relatively slow, because it requires not only an iterative process like PageRank, but also a preprocessing of topical analysis. Gayo-Avello~\cite{Gay13} proposed a more efficient version that does not consider topical analysis, but it has the problem of ranking spammers with high scores. As an alternative to TwitterRank, the {\em InterRank}~\cite{SML13} also uses PageRank (indeed, both measures are highly correlated) on a $\cG_1$ graph with similarity criteria, but it does not require predefined topics.

There also exist extensions of TSPR and TwitterRank. Furthermore, TSPR has been extended to study the semantics of retweet and mention relationships~\cite{WSHC11}. The {\em Topic-Entity PageRank}~\cite{CMC14} looks for influential users by using both topics and entities (e.g., a location, a person, a product, etc.) on a $\cG_2$ graph of retweets. The disadvantage of this measure is that the entities are not always available for all users. Another measure is the {\em TIURank}~\cite{LSML14}. Besides retweets, it also considers the frequency of retweets and strength relationship between users. This strength relationship is estimated by its similarity and a poisson regression-based latent variable model. As it is expected, this measure is also restricted by API limitations. Indeed, it was tested only with small databases. An additional extension is the {\em Author-Reader Influence (ARI)}~\cite{HMR14} model that uses $\cG_2$ graphs, whose edges represent the probability that a tweet posted by a user can be read or shared by another user (by mean of retweets, mentions or replies). The ARI measure considers the influential users as authors of attractive content (relevant and unique), which in theory will be read and replicated by others.

We continue with three additional measures that, like the ARI measure, use PageRank in an efficient way, together with other tools that already exist. The first one is the {\em Twitter user rank}~\cite{NRXT13}, that uses Twitter keyword search (TURKEYS) with retweets, replies and mentions relationships. It is a combination of the Tweet count score activity measure (see Section~\ref{sec:activity}) and the {\em User influence score (UI)}, which is a mechanism based on PageRank that can also use other algorithms such as HITS or SALSA~\cite{LM00}. Alternatively, this measure can also use the tweets dispersion in the network. Then we have the {\em Topic-Sensitive Supervised Random Walks (TS-SRW)}~\cite{KVP15}, that considers a $\cG_2$ graph of mentions and random walks. Unlike the ProfileRank activity measure (see Section~\ref{sec:metric-pagerank}) that also uses random walks, TS-SRW uses similarity of content to support specific topics. Thirdly, we have the {\em Topical Authority}~\cite{HFG13}, calculated on a $\cG_2$ graph of retweets, where the weights of the edges represent the relevance of the content of each retweeted tweet with respect to the chosen topic. This relevance can be determined by any text retrieval model. For their experiments, the authors used the Okapi BM25 ranking function. Additionally, we must mention the {\em Information Amplification Rank} or {\em IARank}~\cite{CS12}, an algorithm slightly weaker than PageRank. In addition to supporting specific events, this is one of the few influence measures that can run in near real time. It corresponds to a linear combination of the FollowerRank popularity measure (see Section~\ref{sec:popularity}) and the {\em Buzz}, given by the number of mentions to the user, divided by its total activity (tweets, retweets and replies) related with a particular event.

In the remainder of this section we include measures that require more complex algorithms. Only the following avoids the Twitter API limitations. For these measures, the algorithms are not explicitly given. The {\em Social Network Influence (SNI)}~\cite{HCA14} is a measure comparable with {\em Klout}, an online influence measure that we shall see in Section~\ref{sec:web}. It is a mixture of several tools. It considers timeline and specific topics, and is calculated from a $\cG_2$ graph of retweets and mentions. Furthermore, it is defined by a formula with adjustable parameters, and considers the probability that a user can influence his followers. In addition, it uses betweenness centrality and PageRank. All in all, it considers the network topology, some activity criteria, and parameters of the user's profile.

Bigonha et al.~\cite{BCMGA12} measured user influence as a combination of centrality, polarity of the users, and quality of their tweets. They used a centrality measure that combines normalized versions of betweenness, eigenvector, in-degree and Twitter Follower-Followee ratio (TFF) on a $\cG_1$ graph and a $\cG_2$ graph of mentions, replies and retweets. Polarity is measured by a sentiment analysis that classifies users into evangelists (those with more positive tweets), detractors (those with more negative tweets) and irrelevants (those with more neutral tweets). Finally, the content quality can be determined by using the {\em Kinkaid factor}~\cite{BCMAG10}, a particular formula based on the number of words, sentences and syllables. Lee et al.~\cite{LL15} also use sentiment analysis, but in this case for the retweeted tweets. With this and the follow-up relationships, the authors can determine the user influence  in terms of their susceptibility and cynicalness.

Sun and Ng~\cite{SN13} constructed a graph whose nodes are tweets, such that the edges are weighted according to explicit relationships (retweets and replies) and implicit relationships (similarity of content). After that, the most influential tweets are measured on different topics using degree centrality, shortest path-cost, and {\em graph entropy}, the latter based on Dehmer~\cite{Deh08}. From the first graph, a second graph of users is extracted, and from here the corresponding most influential users are chosen through algorithms that are able to distinguish between starters and connecters. These algorithms have been applied in online networks with relatively few nodes.

Besides the above, there are more measures that reuse other measures that already exist. The {\em Weighted Ranking Algorithm (WRA)}~\cite{YLLH13} is a linear combination of the {\em ActivityScore} and another measure called {\em QualityScore}, which considers the tweets (filtered by some topic) posted by users that follow another user, the number of tweets of these users, and the probability that tweets are retweeted. WRA is applied on a $\cG_3$ graph of following, tweeting, retweets and replies. Another measure is the {\em Leadership}~\cite{AKRO15}, which is just a linear combination of the {\em Competency} and the {\em Popularity} measures defined before. Finally, the {\em Followship-LDA (FLDA)}~\cite{BTSBC14} model is a slow and particularly complex algorithm that implements a Bayesian Bernoulli-Multinomial mixture model trained by Gibbs sampling. Unlike other measures like TwitterRank, FLDA performs the topical analysis at the same time of computing the influence. It considers offline content analysis and follow-up relationships. Bi et al.~\cite{BTSBC14} also proposed a framework to rank influential users by topic, under different existing measures such as TSPR and TwitterRank. Note that the last two mentioned influence measures do not consider metrics of retweets, mentions nor replies. Actually, they try to rank the referents in a topic, rather than users capable of spreading influence on the network.

We finish with two interesting frameworks. The first one corresponds to R\"abiger and Spiliopoulou~\cite{RS15}. By using supervised learning on specific topics, it was designed to distinguish users according to their influence. The authors do not define a new influence measure. Instead, they observe the real perception of people to classify an actor as influential. They use follow-up relationships, retweets, replies and mentions, as well as the structure and activity network, the centrality of users, and the quality of the tweets. In contrast with the unsupervised environments we have seen above, the authors conclude that for this supervised environment the follow-up relationships are the most important ones to identify influential users. The second framework was design by Kardara et al.~\cite{KPPTV15} to evaluate topical-sensitive measures. They propose a two-dimensional taxonomy classification: by scope (global, local and glocal measures) and by metrics (graphical, contentual and holistic measures). Graphical measures are those that consider topological aspects; contentual measures consider some aspects like content quality, retweets and mentions; and holistic measures consider some aspects shared by the previous two. Furthermore, the authors propose three conditions that a group of real influencers should meet to be compatible with the framework, and two other to determine the relative effectiveness of the measure. This is an interesting attempt to establish an overall classification of influence measures.

\subsection{Predicting influencers}

All the measures seen so far allow ranking the users from known data at a certain instant or period of time. These rankings may change significantly over time, leaving a trail of historical information that can be used to make predictions. To predict influential users, we can extract information from metrics such as the ones given in Table~\ref{tab-metrics}. Xiao et al.~\cite{XZZW13} take advantage of additional information supported by {\em Bitly}\footnote{\url{http://bitly.com}}, referred to the number of short URLs shared by the users, as well as to the number of visits after clicking on them. For most of the remaining measures presented in this section, the details of the algorithms are unknown. In general, these measures have a high complexity, and they use advanced computational tools.

The {\em AWI model}~\cite{YZ12} is a user interaction model that considers the (A)ctivity and (W)illingness of users to retweet through time, in order to measure the (I)nfluence among pairs of users. This model also predicts retweet ratios and influential users. Although the authors apply this model in the context of Sina Microblog, it can also be applied to Twitter. Similarly, the {\em ACQR Framework}~\cite{CXZW13} uses data mining to detect activity (original tweets, retweets and replies), centrality (in-degree and another measure based on Euclidean distance and out-degree), and user reputation (mechanism to distinguish between real users and spammers). It also considers the quality of tweets through the number of replies and retweets, and the reputation of users that reply and retweet. This framework was used to identify and predict the influential users in a relatively small network that was restricted to a specific topic. Soon after, other researchers were able to make online predictions on a small number of user accounts, by combining the ratio of followers and followees with the ratio of tweets written during a certain period of time~\cite{RA14}.

As in the previous section, there also exist predictive influence measures that only use the follow-up relationships, leaving aside the other Twitter actions. Thus, the {\em Time Network Influence Model (TNIM)}~\cite{DJZZHY13} uses a probabilistic generative model to make an offline estimation of the influence power between users. This measure takes into account the time intervals between messages, follow-up relationships, and the relationships of similarity in the content of the tweets. Furthermore, the {\em Author Ranking}~\cite{VRSJLR14} uses the style of the tweets (words, hashtags, websites, references to other accounts) and user behavior (profile information, following ratios, number of tweets, and main user activity, previously determined by a text classification task). This measure considers two steps: A supervised approach and an unsupervised approach, based on Markov Random Field.

In contrast, the following measures take advantage of the different Twitter actions. The {\em ReachBuzzRank}~\cite{SVH13} uses a predictive algorithm based on Hidden Markov Models (HMM) to measure and predict influential users through a statical analysis (based on PageRank) and a temporal analysis (with the new {\em BuzzRank} measure). ReachBuzzRank considers the network topology and the dynamic interactions between users (followers, retweets, replies and mentions). It is restricted to a particular topic, and it was applied on a small network with less than 10,000 nodes. On the other hand, the {\em IDM-CTMP (Information Diffusion Model based on Continuous-Time Markov Process) model}~\cite{LPLSLX14} is a dynamic information propagation model, based on continuous-time Markov process, to predict influential users in a dynamic and continuous way over time. It is applied on graphs whose nodes are users sending tweets, and whose edges relate users that interact (by posting tweets, mentions, retweets or replies) about the same topic. Thus, the paths in the graph generate a succession of tweets over time, sorted by time of emission. The IDM-CTMP model was applied on a database much larger than that used by the ReachBuzzRank. The authors contrast their results with several measures, such as degree and PageRank on $\cG_1$ and $\cG_2$ graphs of mentions and retweets. This measure encourages users who write on various topics, users who are dedicated to spread topics already initiated by others, and users whose tweets are able to generate a rapid response (low latency adoption topic) in other users.

Finally, Bouguessa and Romdhane~\cite{BR15} used a parameterless mixture model-based approach. Firstly, each user is associated with a feature vector that contains information related to their behavior and activity (based on metrics of following, retweets and mentions). After that, through a statistical framework based on multivariate beta mixtures over an offline database, a probability density function is estimated to obtain the users with greater authority.

\subsection{Web applications}
\label{sec:web}

In this section we mention several web applications that put into practice several measures developed with a more applied than academic goal. Although almost all were designed to general purposes, Sameh~\cite{Sam13} described an analytical tool for the study of presidential elections. This application runs in real time and uses data-mining, text-mining, algorithmic graph theory, and sentiment analysis. These techniques have also been used to measure the influence of users on other OSNs~\cite{ZYGQMP14}.

There are websites like {\em Klout}~\cite{RSLD15}, {\em PeerIndex}~\cite{Ser12}, {\em InfluenceTracker}~\cite{RA14}, {\em Twitter Grader}, {\em Favstar}, {\em BehaviorMatrix}, {\em Kred} and {\em Twitalyzer} (besides other that no longer exist, like {\em TurnRank}~\cite{BF11}),\footnote{See \url{http://klout.com}, \url{http://www.peerindex.net}, \url{http://www.influencetracker.com}, \url{https://mokumax.com}, \url{http://favstar.fm}, \url{http://www.behaviormatrix.com}, \url{http://www.kred.com}, \url{http://www.twitalyzer.com}.} that rank the most relevant Twitter users according to activity, popularity, or influence. They use the Twitter API, other OSNs, timeline, and sometimes more complex criteria that are not transparent. At least until 2013, the Klout and PeerIndex measures were well correlated, while the simpler Twitter Grader measures were independent of the previous two. By using machine learning and reverse engineering, we can learn to partially predict the increment of these measure values~\cite{CMT13}. Like previous measures, we must understand that the rankings given by these websites like Klout do not necessarily correspond with the actual influence of the actors in real life~\cite{CDL15,CLD15}.

Most of these applications measure global influencers. A more recent web application is {\em tweetStimuli}~\cite{TSP14},\footnote{\url{http://tweetstimuli.com}} that allow us to rank {\em local influencers}, a concept that comes from Bakshy et al.~\cite{BHMW11}. Given a user $i$, the {\em in-local influence} is calculated with the number of followees of $i$ who have been retweeted at least once by $i$; and the {\em out-local influence} is calculated with the number of followers of $i$ that have retweeted to $i$ at least once. Besides retweets, these definitions also consider tweets marked as favorites or likes. Since the idea is to measure influence at a local level, the offline networks used by tweetStimuli are quite small.

Finally, {\em NavigTweet}~\cite{FH15b}\footnote{\url{http://home.deib.polimi.it/hussain/navigtweet/}} is another application that, among other functions, allows to visualize influential users. It implements a variation of the Analytical hierarchy process (AHP), by considering different metrics as parameters, regarding following, likes, number of tweets, lists, retweets, mentions, hashtags, and URLs mentioned in tweets.

\section{Some additional aspects}
\label{sec:add-aspects}

\subsection{About computational complexity}
\label{sec:comcom}

From a computational complexity point of view, it is more desirable to find algorithms that can run in polynomial time in terms of input size. In algorithmics, the challenge is to build more efficient algorithms in terms of both execution time and the amount of computational memory required. When we consider an input as a data stream, the ``input size'' turns into a less clear concept. The huge amount of information and the high dynamism of Twitter network make a polynomial time algorithm often not enough to capture the current reality of the system, because that reality already have changed. In these cases, if we want to improve the efficiency, the only option is to ignore some of the information we have.

Broadly speaking, there are two types of algorithms to compute measures of influence (or activity, or popularity): Algorithms that run in offline mode and algorithms that run in real time. The algorithms of the first type can leverage more network and user information, and thus to get results more faithful to reality, at the time the data were extracted. In contrast, although the algorithms of the second type cannot use all the information provided by the network, they offer in exchange a quick view of the users status that can be constantly updated.

Almost all measures considered in this article run in offline mode. For instance, all measures based on PageRank and HITS require to know the network graph representation in order to make their calculations. In an offline mode, the biggest problem is whether we use $\cG_2$ or $\cG_3$ graphs, which explicitly represents the high network dynamics. If we consider $\cG_1$ graphs, the problem is the huge amount of Twitter users, together with the Twitter API limitations explained in Section~\ref{sec:API-rest}. For all these cases, the usual practice is to download a large database with Twitter activity within a certain period of time. After that, the required information is filtered and stored in a smaller database that is used as the input to compute the measures. This modus operandi can be very useful in several contexts: If we want to replicate experiments from an already stored database; if we want to study a particular event during a period of time (a natural disaster, a presidential election, etc.); or if both the storing process and the algorithmic execution are efficient enough to yield results relatively frequently (e.g., once daily).

Regarding real-time algorithms, they typically use simple metrics, and avoid working with graphs. Such measures are useful in dynamic websites, because they allow us to display attractive and current network information, e.g., a ranking of the most active current users, or the information spread by some specific set of key users.

The measures designed for Twitter usually do not include a complexity analysis. However, there are some researchers who care about these computational restrictions. For example, some authors are aware of the temporal and spatial complexity of PageRank~\cite{CS12,SGMZ13,HLLC13}. The time complexity of several simple metrics is shown in Table~\ref{tab-metrics} (for a standard textbook related with computational complexity theory, see Garey and Johnson~\cite{GJ79}). Suppose we have a database of Twitter activity over a period of time. In total there are $T$ tweets chronologically ordered. The cost of access to tweets data and metadata in the database is assumed to be constant. Recall that the retweets, mentions and replies are identifiable by patterns described in Section~\ref{sec:twitter}. Since tweets are limited to a length of 140 characters, we also assume that those patterns can be detected in constant time. It is also negligible the cost of comparing two strings, such as user names. Therefore, metrics $OT1$, $OT2$, $OT3$, $RP1$, $RP2$, $RT1$, $RT2$, $FT1$, $FT2$, $M1$ and $M3$ run in $O(T)$ time, because we only need to do a linear search through the list of tweets. Instead, for the metrics $RP3$, $RT3$, $FT3$, $M2$ and $M4$, we also need an auxiliary list to avoid repeated elements. Let be $k$ the size of these auxiliary lists, then these metrics run in $O(T\cdot k)$ time. Metrics $F1$ and $F3$ run in $O(1)$ time, because they just require some simple API requests. Finally, metrics $F2$, $F4$, $F5$ and $F6$ depend on the API limitations presented in Section~\ref{sec:API}.

The time complexity for the different considered measures is shown in Table~\ref{tab:measures}. Several results are obtained directly from the complexity results of the metrics presented above. As metrics $F2$, $F4$, $F5$ and $F6$, those measures that require to know explicit followers or followees are restricted to API limitations. The $F1$ and $F3$ metrics results should be stored ideally when the database is built, because they are subject to future changes. Note that for several topical-sensitive measures we use a lower bound $(\Omega)$ besides an upper bound $(O)$. This is because the upper bound depends on the complexity of the algorithm used to verify whether a tweet belongs to the topic or not.

As we have seen in previous sections, several measures use the PageRank algorithm. Given a graph with $n$ nodes, by using Coppersmith-Winograd method variations for fast matrix multiplication, both PageRank and eigenvalue decomposition can run in $O(n^{2.3727})$ time~\cite{CJB10,Wil12}, and maybe little faster. There are also fast iteration methods like the power method, that allows to compute the dominant eigenpair of a matrix~\cite{BGS05}. Furthermore, given a Twitter database, to build a graph $\cG_2$ or $\cG3$ can be done in $O(T)$ time. That is why measures like Starrank, TrueTop and others can run in $O(T+n^{2.3727})$ time. H-index only takes $O(T+n)$ time, because it is well known that given a graph, this measure can be computed in linear time. Measures like the Acquaintance-Affinity Score and the Action-Reaction run in $O(T\cdot k^2)$ time because they require an auxiliary matrix instead of an auxiliary vector. The time complexity for some complex measures remains unknown.

\subsection{About correlation of measures}
\label{subsec:corr}

The centrality measures are usually correlated with each other to determine similarities or dependencies between them~\cite{Bol88,RPWDMK95,Fau97,VF98}. Two similar measures provide similar rankings and thus might be redundant, while two very different measures provide different classification criteria and may be useful for different purposes. Altough traditional measures like degree, betweenness, closeness and eigenvector are conceptually related, from a correlation analysis it can be shown that they are different~\cite{VCLC08}. The correlation mechanisms come from statistics, and are widely used in centrality measures for general networking.

In the context of Twitter, researchers often use, reasonably, the more traditional correlation coefficients, namely the {\em Spearman's rank correlation coefficient} $(\rho)$~\cite{Spe04} and the {\em Kendall Tau rank correlation coefficient} $(\tau)$~\cite{Ken38}. Let $w$ and $y$ be two lists of $n$ users each, we have~\cite{YW13}:
$$\rho=1-\frac{6\sum(x_i-y_i)^2}{n(n^2-1)} \mbox{ \, and \, } \tau=\frac{n_c-n_d}{0.5n(n-1)}$$
where $x_i$ and $y_i$ are the rankings of the users $i$ in the lists $x$ and $y$, respectively. Furthermore, $n_c$ is the number of concordant pairs $(i,j)$ (i.e., such that either $x_i>x_j$ and $y_i>y_j$, or $x_i<x_j$ and $y_i<y_j$) and $n_d$ is the number of discordant pairs, i.e., those that are not concordant.

The values of both $\rho$ and $\tau$ are in the $[-1,1]$ interval, where $1$ means that both measures are equal, $0$ that they are completely independent, and $-1$ that one is the inverse of the other. In our context, where we can have thousand and even millions of users, it is usual to compare only the first hundreds or thousand places of the rankings. This usually generates two lists with different users. To solve this we can compute the following:
$$\frac{\rho_k(x,y)+\rho_k(y,x)}{2} \mbox{ \, and \, } \frac{\tau_k(x,y)+\tau_k(y,x)}{2}$$
where $\rho_k(x,y)$ and $\tau_k(x,y)$ are obtained by matching with the following steps~\cite{YW13}:

\begin{enumerate}
 \item Choose the first $k$ users of list $x$.
 \item Determine the rankings in list $y$, establishing a new list $y'$ that only ranks the users considered in $x$.
 \item Correlate the first $k$ users of $x$ with those of $y'$.
\end{enumerate}

Other known coefficients are the {\em Pearson correlation coefficient}~\cite{Pea95} and the {\em Goodman-Kruksal gamma}~\cite{GK54}. If the rankings are known in advance (e.g., through a system of surveys) we can use {\em precision and recall}, two measures used in information retrieval for evaluating systems where users have a binary classification (relevant / not relevant, influential / not influential, etc.)~\cite{CS12}.

\subsection{Some experiments}
\label{sec:experiments}

In this section, we present the results of a few number of measures on the same network, in order to show how the rankings change depending on the different considered centrality criteria. The network dataset is provided by the Stanford Large Network Dataset Collection (SNAP)~\cite{LK14}. 

The dataset represents all the tweets related with the Higgs boson experiment that were posted between 1st and 7th July 2012. It allows to generate a $\cG_1$ graph with 456,626 users nodes, and it also contains interactions of retweets, replies and mentions between the users~\cite{DLMM13}. With this data the following metrics were computed: $RP1$, $RP3$, $RT1$, $RT2$, $RT3$, $M1$, $M2$, $M3$, $M4$, $F1$ and $F3$.

\begin{figure}[t]%
\centering
\includegraphics[width=\textwidth]{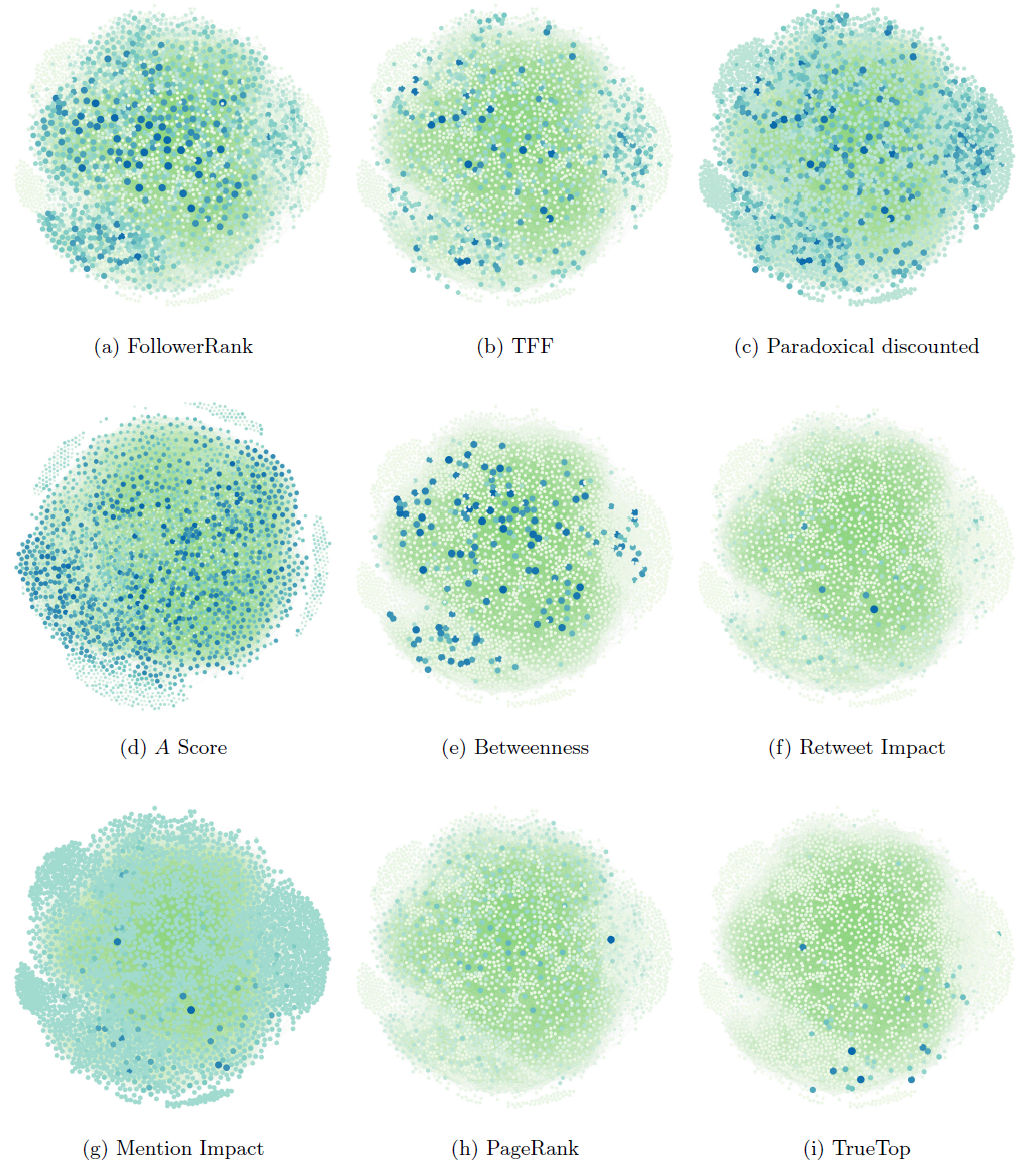}
\caption{Different measures computed on the same Twitter network. The most central (popular or influential) user nodes are larger and more blue, while the less central ones are smaller and more white.}
\label{fig:measures}
\end{figure}

We computed nine measures (four of popularity, and five of influence) that have been mentioned in this work. The activity measures could not be computed because the dataset does not provides activity information like the total number of tweets. To illustrate each measure, we created a $\cG_1$ subgraph from the union of the thousand top nodes of each measure. This results in a subgraph with 2842 nodes and 62637 edges. The different measures are illustrated in Figure \ref{fig:measures}. Note that some measures that are not normalized present a few nodes with values significantly higher than others. Furthermore, it seems from the figures that the FollowerRank is very similar to the $A$ Score, and the TFF is very similar to the Paradoxical discounted. This kind of conclusions can be partially explained from their respective formulas. However, a more formal demonstration requires the use of correlation coefficients. We leave this correlation analysis as future work.

\section{Open research problems}
\label{sec:open}

Identifying the influential actors on Twitter is a research under continuous development. The check marks in Table~\ref{tab-metrics} put in evidence the open problem of defining a measure covering all types of Twitter metrics. In this sense, there are some measures easily generalizable to metrics still not considered. For example, it does not seem complicated to generalize the SNP measure, in order to include also the favorites or likes. These new measures, of course, must continue to be efficient, in the context of Section~\ref{sec:comcom}.

Despite of the above, there are already many measures, and new ones are constantly appearing. We believe the problem is not necessarily creating new influence measures (unless they really provide new ranking criteria, so far not considered), but classifying these measures so that we can choose from the most useful for our purposes. In this sense, the classification presented in Table~\ref{tab-metrics} could still be improved. Note from the question marks in the chart that the time complexity for some measures remains unknown. Furthermore, new classification criteria could be added, such as the ones suggested by Kardara et al.~\cite{KPPTV15} for general topical-sensitive measures.

In the same vein, a very interesting study that we leave open is to comprehensively correlate all measures contained in this paper, in order to identify similarities and differences, redundancies and independencies. Note that the similarity of check marks in Table~\ref{tab-metrics} does not represent a correlation between the measures. Indeed, as we have seen, there is no direct correlation between active, popular and influential actors~\cite{RGAH11}. Furthermore, there exist measures that use several kind of metrics, which are well correlated with other measures that only use a strict subset of those metrics. Such is the case of the AAI measure, for instance, that is well correlated with the number of followers~\cite{SST13}. The experiments of Section~\ref{sec:experiments} suggest that the FollowerRank and the $A$ Score could be another example.

In addition, there are other ways to detect influential users without the use of centrality measures. For instance, Morone and Maske~\cite{MM15} study the problem of finding the minimal set of nodes (which they called the {\em optimal influencers}) necessary to spread information to the whole network. Removing the optimal influencers from the network leads to its collapse. The authors propose to solve this problem by solving the mathematical problem of finding an optimal percolation. They conclude that under this criterion, for $\cG_1$ graphs, the influencers are not usually users with too much followers. It would be interesting to know what happens if we consider $\cG_2$ and $\cG_3$ graphs.

Finally, besides influential users, there are several measures that rank the most influential {\em tweets} in certain period of time. A survey on research of these kind of measures could also be performed. In that case, we must consider some relevant aspects, such as the content quality of tweets, or the influence exerted by a tweet over the readers~\cite{NNH13}.

\section{Conclusions}
\label{sec:conclusions}

In this survey we have classified the various measures of activity, popularity and influence in the literature for the specific context of Twitter. As far as we know, this is the first comprehensive survey of this type. The main results can be seen in Table~\ref{tab:measures}. The diversity of measures is very high, especially considering that the oldest ones were defined only in 2009. From this chart we can see that almost all popularity measures are associated with follow-up relationships, while most activity measures consider actions of replies. In turn, for influence measures we highlight the use of retweets. Of all the kinds of metrics considered in Table~\ref{tab-metrics}, the least used are related to the favorites or likes. This may be because this functionality took more years to be implemented and be known by Twitter users. Despite of this, the new measures should begin to consider such interactions, which are so well supported by the Twitter API as the others.

The ckeck marks in Table~\ref{tab-metrics} do not necessarily reflect the quality of the measures. They just realize the variety of considered criteria. Note that no current measure uses all the different kind of metrics. The few measures that use metrics of favorites (or likes), always dispense with some another kind of metric. On the other hand, we have seen that using metrics like $F2$, $F4$, $F5$ and $F6$ considerably increases the time response, because they require a precomputing period to provide implicit follow-up relationships among the actors. These metrics can be useful to study particular events, when data is collected in a complete and fixed database that does not need to be updated. Otherwise, it is best to avoid them.

Interestingly, almost one half of the existing influence measures are based in one way or another on the PageRank algorithm. This algorithm, originally created to assess the relevance of websites on the Internet, has been settled as a powerful tool to rank influential users in online social networks. Several efforts have also been devoted to the definition of topical-sensitive measures, which must necessarily include a content analysis of the published tweets. Instead, predictive influence measures are few. Some algorithms for these measures are not explicitly given, so they cannot be used by other researchers.

We hope this survey will help researchers and developers to know the variety of centrality measures for Twitter network. This variety of criteria to identify influential actors is a proof that the concepts related with the spread of influence in social networks have not yet reached a consensus.

\section*{Acknowledgements}
This work was partially funded by project PMI USA1204.\\
We thank the anonymous referees for their comments and suggestions that helped us to improve the content and the presentation of the paper.

\clearpage


\end{document}